\documentclass[pra,twocolumn,superscriptaddress,showpacs,amsmath,amssymb]{revtex4-1}
\usepackage{latexsym}

\usepackage{amsmath, amsthm, amssymb} 

\usepackage{txfonts}

\usepackage{comment}

\usepackage[latin1]{inputenc}
\usepackage{amsmath}
\usepackage{amsfonts}
\usepackage{amssymb}
\usepackage{graphics}
\usepackage{setspace}
\usepackage[usenames]{color}
\usepackage{array}
\usepackage{natbib}
\usepackage{epsfig}
\usepackage{graphicx}
\usepackage{amsmath}
\usepackage{amsfonts}
\usepackage{amssymb}
\usepackage{cancel}
\usepackage{graphicx}
\usepackage{ulem}
\usepackage{wrapfig}
\usepackage{epsfig}
\usepackage{graphicx}
\usepackage{amsmath}
\usepackage{amsfonts}
\usepackage{amssymb}
\usepackage{graphicx}
\usepackage{wrapfig}
\usepackage{textcomp}
\usepackage{filecontents}
\usepackage{pgf,pgfarrows,pgfnodes,pgfautomata,pgfheaps,pgfshade}
\usepackage[colorlinks=true, linkcolor=blue]{hyperref}
\usepackage{lipsum}
\usepackage{balance}
\usepackage[latin1]{inputenc}
\usepackage{blindtext}
\usepackage{textgreek}

\makeatletter
\def\maketitle{
\@author@finish
\title@column\titleblock@produce
\suppressfloats[t]}
\makeatother

\begin{document}

\title{
Cavity-QED of a quantum  metamaterial with tunable disorder
}

\author{Grigoriy~S.~Mazhorin
}
\affiliation{
Moscow Institute of Physics and Technology, Dolgoprudny, 141701, Russia}
\affiliation{
National University of Science and Technology MISiS,   119049 Moscow, Russia}
\affiliation{
Russian Quantum Center,   Skolkovo, 143025 Moscow Region, Russia}

\author{Ilya~N.~Moskalenko
}
\affiliation{
National University of Science and Technology MISiS,   119049 Moscow, Russia}
\affiliation{
Russian Quantum Center,   Skolkovo, 143025 Moscow Region, Russia}

\author{Ilya~S.~Besedin
}
\affiliation{
National University of Science and Technology MISiS,   119049 Moscow, Russia}
\affiliation{
Russian Quantum Center,   Skolkovo, 143025 Moscow Region, Russia}

\author{Dmitriy~S.~Shapiro
}
\email{dmitrii.shapiro@kit.edu} 
\affiliation{
Dukhov Research Institute of Automatics (VNIIA),  Moscow 127055, Russia}
\affiliation{
V. A. Kotel'nikov Institute of Radio Engineering and Electronics, Russian Academy of Sciences, Moscow 125009, Russia}
\affiliation{
Institute for Quantum Materials and Technologies, Karlsruhe Institute of Technology, 76021 Karlsruhe, Germany}

\author{Sergey~V.~Remizov
}
\affiliation{
Dukhov Research Institute of Automatics (VNIIA),  Moscow 127055, Russia}
\affiliation{
V. A. Kotel'nikov Institute of Radio Engineering and Electronics, Russian Academy of Sciences, Moscow 125009, Russia}
\affiliation{
Department of Physics, National Research University Higher School of Economics, Moscow 101000, Russia}

\author{Walter~V.~Pogosov
}
\affiliation{
Dukhov Research Institute of Automatics (VNIIA),  Moscow 127055, Russia}
\affiliation{
Institute for Theoretical and Applied Electrodynamics, Russian Academy of
Sciences, 125412 Moscow, Russia}
\affiliation{
HSE University, 109028 Moscow, Russia}

\author{Dmitry~O.~Moskalev
}
\affiliation{
Dukhov Research Institute of Automatics (VNIIA),  Moscow 127055, Russia}
\affiliation{
FMN Laboratory, Bauman Moscow State Technical University, Moscow 105005, Russia}

 \author{Anastasia~A.~Pishchimova
 }
 \affiliation{
 Dukhov Research Institute of Automatics (VNIIA),  Moscow 127055, Russia}
\affiliation{
FMN Laboratory, Bauman Moscow State Technical University, Moscow 105005, Russia}

\author{Alina~A.~Dobronosova
}
\affiliation{
Dukhov Research Institute of Automatics (VNIIA),  Moscow 127055, Russia}
\affiliation{
FMN Laboratory, Bauman Moscow State Technical University, Moscow 105005, Russia}

 \author{I.~A.~Rodionov
 }
 \affiliation{
 Dukhov Research Institute of Automatics (VNIIA),  Moscow 127055, Russia}
\affiliation{
FMN Laboratory, Bauman Moscow State Technical University, Moscow 105005, Russia}

\author{Alexey~V.~Ustinov
}
\affiliation{
National University of Science and Technology MISiS,   119049 Moscow, Russia}
\affiliation{
Russian Quantum Center,   Skolkovo, 143025 Moscow Region, Russia}
\affiliation{
Physikalisches Institut, Karlsruhe Institute of Technology, 76131 Karlsruhe, Germany}

\begin{abstract}
We  explore experimentally a quantum metamaterial based on a superconducting chip with 25 frequency-tunable transmon qubits coupled to a common coplanar resonator. The collective bright and dark modes are probed via the microwave response,  i.e., by measuring the transmission amplitude of an external microwave signal.
All qubits have individual control and readout lines. Their frequency tunability allows to change the number $N$ of resonantly coupled qubits and also to introduce a disorder in their excitation frequencies with preassigned distributions.  While increasing $N$, we demonstrate the expected $N^{1/2}$ scaling law for the energy gap (Rabi splitting) between bright modes around the cavity frequency. By introducing a controllable disorder and averaging the transmission amplitude over a large number of realizations, we  demonstrate a decay of mesoscopic fluctuations which mimics an approach towards the thermodynamic limit. The collective bright states survive in the presence of disorder when the strength of individual qubit coupling to the cavity dominates over the disorder strength.

\end{abstract}

\maketitle

\section{Introduction}
During last years, superconducting  qubits have shown remarkable progress  in realizations of scalable  quantum computing devices~\cite{Arute:2019aa} as well as in fundamental studies of circuit quantum electrodynamics (QED)~\cite{Clerk:2020aa}.
Quantum circuits based on superconducting  qubits  allow for testing  fermion models~\cite{Barends:2015aa}, geometric phases~\cite{PhysRevLett.113.050402}, weak localization~\cite{Chen:2014aa}, topologically ordered states~\cite{Roushan:2014aa,besedin2021,SSH-Fernandez-Rossier}, and beyond.
Various phenomena related to photonic transport and photon-photon interaction  can be observed even for a circuit with a single qubit. They appear when microwave photons are transmitted through a qubit circuit which plays a role of a  nonlinear oscillator. Examples are photon blockade~\cite{PhysRevLett.107.053602,Photon_Blockade_corr}, transfer of thermalized  photons  and measurement of their bunching~\cite{Goetz:2017aa},  probing of transmitted photons  statistics~\cite{Dmitriev:2019aa,Honigl-Decrinis:2020aa,Zhou:2020aa},   
 and multi-photon transitions~\cite{Braumueller2015}.  

Multi-qubit circuits find their applications in  quantum metamaterials~\cite{macha2014implementation}, which are examples of  artificial quantum matter with tunable properties. The dynamics of such metamaterials  is governed by  quantum-optical models, such as  Dicke~\cite{Frisk-Kockum:2019aa,kirton2018introduction,Shapiro2020} or Bose-Hubbard~\cite{Biella:2015aa,Vicentini:2018aa,Gleb-PRL-2021} models, which capture the physics of coupled photonic modes  and qubit degrees of freedom. 

The major technical challenge for fabrication of multi-qubit metamaterials is in making the energy level separations $h\epsilon_j$ of many non-identical qubits as similar as possible. This is required for observing, e.g., a coherent response of the metamaterial and collective bright modes of the system. It has been argued~\cite{macha2014implementation,Shapiro2015,Shulga2017} 
but not yet proved that, for non-tunable qubits, this problem can be overcome by engineering large enough qubit coupling strength $g$ to the cavity, similar to the way of overcoming the effects of inhomogeneous broadening in lasers made of natural atoms. Coherent response of a metamaterial can be expected if the spread in $\epsilon_j$ becomes smaller than $g$. The individual qubit frequencies  $\epsilon_j$ can be individually controlled by applying local fields to qubits, which, obviously, becomes more and more technically difficult when increasing the number of qubits in a metamaterial. 

Multi-qubit metamaterials represent a mesoscopic limit of naturally occurring ensembles consisting of nominally identical atoms or spins. Here, however, the fluctuations are different for each atom and lead to the resonance line broadening in the presence of fluctuating local fields and interactions between atoms. The homogeneously broadened emission (lifetime-limited) line has a Lorentzian profile, while the inhomogeneously broadened emission will have a Gaussian profile. While fluctuations in the thermodynamic limit corresponding to a very large number of emitters are well understood and studied in solid-state and molecular spectroscopy, the mesoscopic limit of a countable (not too large) number of emitters is very difficult to explore with natural atoms or spins. Qubit metamaterials may be suitable to fill this knowledge gap. One of the prominent examples is a spin ensemble coupled to a cavity and described by the Tavis-Cummings model~\cite{Tav-Cumm-1968}. 
Here, qubits with individual frequency control can be used to introduce a tunable static or dynamic disorder.

In this work, we report on experimental realization of a multi-qubit platform that allows to simulate disorder effects in quantum metamaterials. We have designed and fabricated a superconducting chip based on an array of 25 transmon~\cite{PhysRevA.76.042319} qubits coupled to a common coplanar resonator. The excitation frequency of every qubit is individually tunable in GHz range. Hence, arbitrary disorder realizations can be easily implemented and studied in this setting.

 Our interest in disordered quantum metamaterials is twofold. On one hand, disorder in the frequency of emitters coupled to a cavity is an important technical issue in devices that rely on coherent operation. In the weak coupling limit, these effects result in inhomogeneous broadening. Simulators based on superconducting qubits are no exception to this issue. Digital simulation approaches based on Trotterization~\cite{Smith:2019aa} circumvent this issue by using well-calibrated gates to approximate the evolution of a system under a continuous-time Hamiltonian. However, the tradeoff is that even relatively simple simulations require a large number of Trotterization steps, and the increase in the amount of these steps results in a rapidly decaying simulation fidelity. For analogue simulators, the disorder in the qubit frequencies stems from the varying critical currents of the Josephson junctions. Correcting for this frequency using SQUIDs moves the qubits away from their flux sweet spots, which significantly degrades their coherence properties~\cite{PhysRevApplied.13.054079}. Thus, for analogue simulation, disorder and control infidelity is the major source of errors in the simulation.
For digital simulations, various randomization-based techniques have been developed to enhance signal~ \cite{arute2020observation}. Here we use randomization to investigate emergent dark states in a system of transmon qubits coupled to a common microwave cavity.

On the other hand, our interest in disordered quantum metamaterials  is   motivated by  theoretical studies~\cite{PhysRevLett.114.196402,PhysRevLett.114.196403,PhysRevB.102.144202} where an intriguing interplay between a coherent collective coupling and disorder was discussed. As shown in these studies, the structure of eigenstates has a strong impact on photon transmission.  Photon transport measurements   allow to distinguish localized or semilocalized regimes which exhibit either exponential or  power-law decays  of transmission amplitude with $N$, respectively. In the semilocalized case, wave functions of dark states are neither localized nor extended~\cite{PhysRevB.102.144202}.
We also mention a recent analysis of transmon-based quantum computing  networks~\cite{berke2020transmon} which are systems with  built-in differences in physical qubit  parameters. As shown in~\cite{berke2020transmon}, physics of disordered spin ensembles, in particular many-body localization, becomes crucial for an operation of those systems.

\section{Theoretical background}
We address the low excitation regime, where the rotating-wave approximation is valid and the Dicke model for $N$ qubits is reduced to the     Tavis-Cummings model. For   particular     qubit frequencies $\epsilon_j$,
qubit-cavity couplings $g_j$, and  the cavity mode with the frequency $\nu_c$, the Tavis-Cummings model  reads as 
 \begin{equation}
 	\hat H =\nu_c \hat a^\dagger \hat a + \sum\limits_{j=1}^N \epsilon_j \hat\sigma_j^+\hat\sigma_j^- + \sum\limits_{j=1}^N g_j( \hat\sigma_j^+ \hat a+ \hat a ^ \dagger \hat\sigma_j^-) \ . \label{H}
 \end{equation}
 Here, $\hat a ^\dagger$ and $\hat a$ are the photon creation and annihilation operators,  $ \hat\sigma_j^+$ and $\hat\sigma_j^-$ are raising and lowering Pauli operators acting upon $j$-th two-level system. 
 In the low energy limit, this Hamiltonian can be represented as a $N+1$-dimensional matrix $\mathcal{H}_{i,j}=\langle \psi_i | \hat H |\psi_j\rangle$  after the projection of $\hat H$ on a single excitation basis, $\{|\psi_i\rangle\}_{i=1}^{N+1}= \{ \hat a^\dagger |g.s.\rangle; \ \hat \sigma_1^+ |g.s.\rangle; \ ... ;\  \  \hat\sigma_N^+ |g.s.\rangle \}$.

 An analysis of  the disordered   Tavis-Cummings model  
is complicated  because the bright polariton modes are not decoupled from   dark  states   anymore~\cite{kirton2018introduction}, and the relevant  Hilbert space is enlarged. This results in such phenomena in inhomogeneously broadened systems as a competition between superradiance and    dephasing~\cite{PhysRevLett.95.243602}, and cavity protection effect~\cite{PhysRevA.84.063810}.

 The respective Green function matrix that takes into account a dissipation to an environment, is   $\mathcal{G}_{\rm  }(\omega)=(\omega \mathcal{I} +i\mathcal{D}-\mathcal{H})^{-1} \ .$   Here, $\mathcal{I}=\delta_{i,j}$ is the identity matrix in the   basis $\{|\psi_i\rangle\}_{i=1}^{N+1}$   and the matrix $\mathcal{D}={\rm diag} [\kappa, \Gamma_1, ... , \Gamma_N]$ is determined by the loss rate  in the resonator, $\kappa$, and the relaxation from the excited to the ground state in qubits, $\Gamma_i$. 
It can be written through
 the   Green functions of decoupled  resonator and qubits, $G_{\rm ph}(\omega)=(\omega+i\kappa-\nu_c)^{-1}$ and $G_{{\rm q},j}(\omega)=  (\omega+i\Gamma_j-\epsilon_j)^{-1}$, respectively, as 
$
 	\mathcal{G}_{\rm  }(\omega)=\begin{bmatrix}
 		G_{\rm ph}^{-1}(\omega) && -\mathbf{g}^T \\
 		-\mathbf{g} && \mathbf{G}_{{\rm q}}^{-1}(\omega)
 	\end{bmatrix}^{-1} 
 $.
Here the $N$-dimensional matrix $\mathbf{G}_{{\rm q}}(\omega)= \delta_{i,j}G_{{\rm q},j}(\omega)$ and the vector $ \mathbf{g}=(g_1; \ g_2; \ ...; \ g_N)^T$ with all coupling constants are introduced. Photonic propagator   $\mathcal{G}_{\rm  ph}(\omega)\equiv [\mathcal{G}_{\rm  }(\omega)]_{1,1}$ is found after an expansion of $\mathcal{G}_{\rm  }$ by the non-diagonal part   and resummation of the first diagonal element. The result is
\begin{equation}
	\mathcal{G}_{\rm  ph}(\omega)=\frac{1}{G_{\rm ph}^{-1}(\omega)-\mathbf{g}^T \mathbf{G}_{{\rm q}}(\omega) \mathbf{g}}   \ , \label{G-ph}
\end{equation} 
The  self-energy   term   $ \mathbf{g}^T \mathbf{G}_{{\rm q}}(\omega) \mathbf{g}=\sum\limits_{j=1}^N\frac{g_j^2}{\omega+i\Gamma_j-\epsilon_j}$ takes into account the diagonal disorder in $\epsilon_j$ and  non-diagonal disorder in $g_j$.  This Green function approach is in    agreement with earlier work~\cite{PhysRevLett.53.1732} where a solution  for a susceptibility  has been found from the master equation for a density matrix.

Resolving the equation $\mathcal{G}_{\rm ph}^{-1}(\omega)=0$ with respect to $\omega$  in the absence of the disorder,  $\epsilon_j=\epsilon$, one finds that the frequencies  of bright collective modes are $\nu_\pm =\frac{1}{2}\left(\nu_c+\epsilon\pm \sqrt{(\nu_c-\epsilon)^2+4|\mathbf{g}|^2} \right) $. If   $g_i=g$ and resonant condition holds, $\epsilon_j=\nu_c$,  then one finds a well-known scaling  of the energy gap with $N$ in the Tavis-Cummings model, $\nu_+ - \nu_- =2g\sqrt N$.
The initial task in this work is thus to demonstrate this  scaling law $\propto N^{1/2}$. It becomes possible by means of subsequent increase of qubit number tuned into the resonance with the photon mode.

The main aim of this work is to study an ensemble of qubits with tunable  diagonal  disorder. We set our goals to demonstrate the effect of self-averaging in transmission amplitudes of disordered ensemble and to observe mesoscopic fluctuations which decrease with $N$. As long as the diagonal disorder is fully controllable, we set  the resonant condition between the resonator mode and all qubits on the average as $\nu_c=\langle \epsilon_j\rangle $. The   probability density to find the $j$-th qubit in a frequency range $[\epsilon; \epsilon+d\epsilon]$ is simulated by a  flat   function $
p(\epsilon )=\frac{1}{\Delta}\theta(\Delta/2-|\epsilon -\nu_c|) $ which is symmetric near $\epsilon=\nu_c$  and has  a controllable spread $\Delta$.  The non-diagonal disorder effects are  less interesting. In our regime of low excitation numbers     it results in a  renormalization of the effective coupling. This can be seen from   the self-energy term where $g_i$ appear in numerators. Oppositely,  $\epsilon_j$ appear in denominators   and the averaging by this variable becomes more non-trivial. 

To explore the mesoscopic effects in the qubit metamaterial, we collect data for the transmission  coefficient  $S_{21}$ of the microwave  probe signal sent at  the cavity bare   frequency $\omega=\nu_c$. As follows from in-out theory where a matching of reflected and transmitted waves is preformed, the transmission coefficient is related to the Green function  (\ref{G-ph}) as follows,
$
S_{21} = \sqrt{\gamma_{\rm in} \gamma_{\rm out} }\mathcal{G}_{\rm ph}(\nu_c)  
$.
Here, $\gamma_{\rm in}$ and $\gamma_{\rm out}$ are   radiation rates  from the resonator into {\it in}- and {\it out}-waveguides. 

Let us analyze   $\langle S_{21} \rangle $ where the averaging is based on a large number of  diagonal disorder realizations with the probability density $p(\epsilon)$. We consider first the limit of large   $N$ where, similarly to the thermodynamic limit, fluctuations are suppressed and self-averaging can be applied.  Namely, the sum  from  (\ref{G-ph})  is replaced by the integral with $p(\epsilon)$ which is    the density of states analogue. One finds $\mathbf{g}^T \mathbf{G}_{{\rm q}}(\omega) \mathbf{g}=\pi g^2 N/\Delta$ and, consequently, the transmission coefficient  is
\begin{equation}
	\langle S_{21}  \rangle=\sqrt{\gamma_{\rm in} \gamma_{\rm out} } \frac{-i}{\kappa + \pi g^2  N/\Delta  } \ . \label{S12-aver}
\end{equation}
 The detailed derivation is presented in Appendix~\ref{App:A}.
Mesoscopic fluctuations $ \Delta S_{21}    \equiv  S_{21}- \langle S_{21}   \rangle$, which are considered as a random value corresponding to a particular disorder realization, appear at finite $N$.
They are   found  as follows in the leading order by the finite $N$
\begin{equation}
\langle |\Delta S_{21}|^2 \rangle=\gamma_{\rm in} \gamma_{\rm out}\frac{\pi N g^4}{2\Gamma \Delta(\kappa + \pi g^2  N/\Delta)^4 }  \ .  \label{deltaS12^2}
\end{equation}
 One sees that the average decays as $ \langle S_{21}  \rangle\propto N^{-1}$ while  fluctuations   as $\propto N^{-3}$ at large $N$ limit.
In our experiment, we use  these results (\ref{S12-aver}) and  (\ref{deltaS12^2}) to fit the measured data.

\section{Quantum circuit and measurement setup}
\begin{figure}[h!]
	\center{\includegraphics[width=\linewidth]{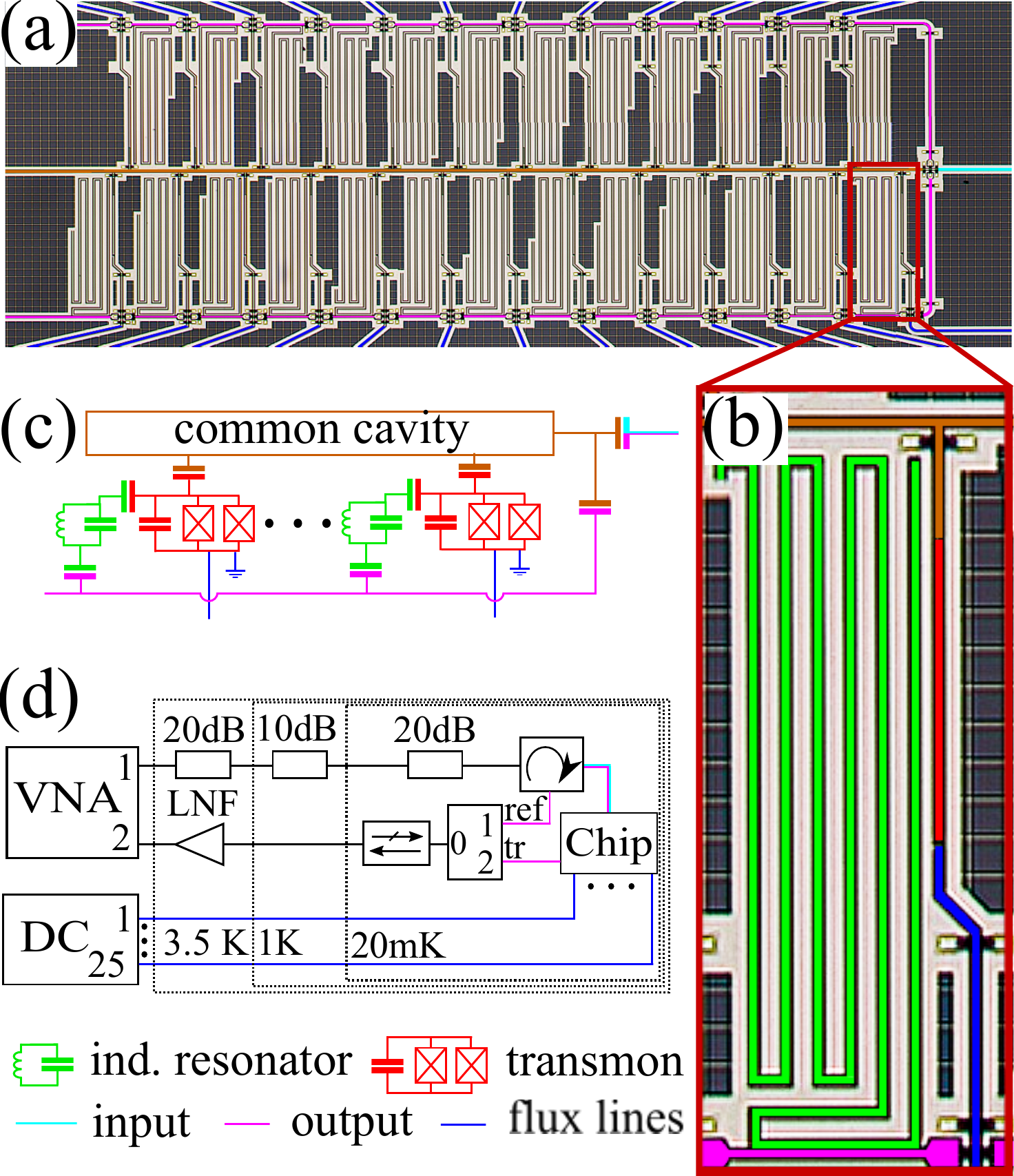}}
	\caption{Superconducting circuit  and scheme of measurement. {\bf (a)} False-colored optical photograph of the quantum metamaterial implemented as a chip with 25 superconducting qubits (transmons). {\bf  (b)} Enlarged fragment of the setup showing a single qubit, its individual readout resonator, flux bias line and a fragment of the common cavity. {\bf (c)} Equivalent electric circuit of the device. {\bf (d)} Setting of the microwave measurement.}
	\label{figScheme}
\end{figure}

Here we present first our experimental  setting which involves 25 superconducting transmon qubits. A particular qubit, measurement scheme, and equivalent circuit are shown in Figs.~\ref{figScheme}~(a-d), respectively.
Qubits are   capacitively coupled to a common cavity  realized as a  $\lambda/2$ coplanar waveguide resonator (panel~(a), brown color). The cavity is terminated with the input transmission line    (cyan color). A transmon qubit   (panel~(b)), at  one of its   ends, involves a short  segment running close to the common cavity conductor (brown color).  
    This results in a strong capacitive qubit-cavity coupling. At the  other end, the transmon is connected to an individual flux bias  line (blue color) that allows to tune the excitation frequency of a particular qubit. There are 25 individual  control lines in total; they are implemented as  asymmetric  SQUIDs connected to a  multi-channel DC source.  Fabrication steps of the device are described in Appendix~\ref{Sample fabrication}.
 Each of the qubits is capacitively coupled to an individual readout resonator  
(panel~(b), green color). These resonators have different lengths and frequencies and operate in a dispersive regime. This allows us to address each of the qubits individually through this feed line using frequency multiplexing, perform two-tone spectroscopy of qubits, and perform calibration of the frequency controls (Appendix~\ref{Calibration}).

The common cavity is connected to its own microwave feedline in a butt-port geometry.
Readout resonators are notch-port coupled to a common cavity and { are} connected to the  output line (magenta color in   the circuit shown in panel~(d)).

We perform measurements of the reflection and transmission amplitudes, $S_{11}$ and $S_{21}$, with the use of a microwave circulator and  a switch. In  $S_{11}$ measurements, the incident signal is sent to the  common cavity and reflected back.
In $S_{21}$ measurements, the incident signal excites qubit modes and leaves the chip through the individual resonators coupled to qubits. 
    
 The   measurement scheme  is shown in Fig.~\ref{figScheme}~(c). 	The microwave drive tone, sent from the vector network analyzer (VNA), is attenuated by 50~dB before entering the chip. After passing through the chip, the signal is amplified and measured by the VNA, yielding the complex transmission amplitude $S_{21}$. Due to long attenuation and amplification chains,    $ S_{21} $ is 
not calibrated. Thus, the data are presented relative to an arbitrary level hereafter. 
 		
 The spectroscopy data for  $S_{21}$, where frequencies of  resonant qubits, $\epsilon=\epsilon_j$, and probe signal, $f_p$,  vary, are shown in Figs~\ref{figVRS}~(a-d).  Here, we present results for $N=4,7,16,21$ resonant qubits. These measurements are performed using the specific calibration procedure (Appendix~\ref{Calibration}).
   Bright anticrossings marked by black dashed curves are the energies of  Rabi collective modes $\nu_\pm$. Yellow and red dashed lines are bare frequencies of the cavity and resonant qubits. The increase of the gap with $N$ indicates for a bright-state coherence between qubit array  and photon mode.

\section{Vacuum Rabi splitting}
\begin{figure}[h!]
	\center{\includegraphics[width=\linewidth]{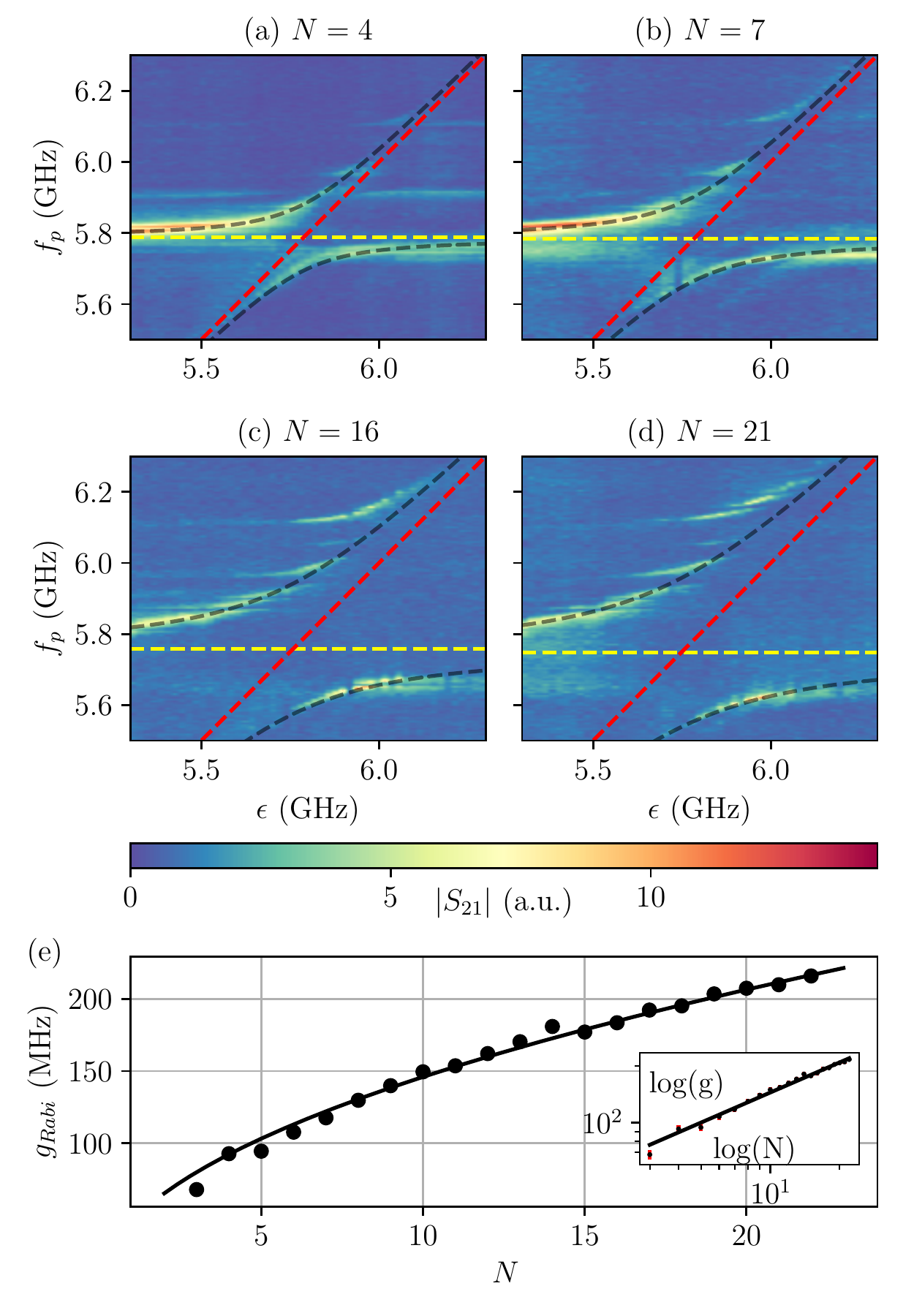}}
	\caption{Spectrum of collective bright (Rabi) modes.
		{\bf (a)}-{\bf (d)} Reflection amplitude $|S_{21}|$ measured for different numbers of tunable qubits ($N=4,7,16,21$). Black dashed lines are $\nu_\pm$ energies of Rabi satellites predicted by the Tavis-Cummings   model.
		Yellow dashed lines stand for bare cavity mode frequency $\nu_c$. Red dashed lines are bare frequencies of tunable qubits. {\bf (e)} $N^{1/2}$-scaling of  Rabi splittings $g_{\rm Rabi}$ for $N=3, ... \ , \ 23$ qubits tuned into the resonance with the cavity.  Inset: the data in logarithmic coordinates   fitted  by  $g_{\rm Rabi}=g N^{\alpha}$  (solid line). }
	\label{figVRS}
\end{figure}

The first important result of this work is the demonstration of $N^{1/2}$-scaling in the Rabi splitting, $g_{\rm Rabi}$, as a function of  $N$, which is the number of qubits tuned into the resonance with the cavity.
In Fig.~\ref{figVRS}~(d)   the dependence of  $g_{\rm Rabi}(N)$,  where $N$ changes from 3 to 23, is demonstrated.   The  theoretical dependence  $g_{\rm Rabi}= g \sqrt{N}$  (solid curve) shows good agreement with the  experimental data (dots).  In the inset these data are shown in  logarithmic    coordinates. The points are    approximated  by  $ g_{\rm Rabi}= g N^{\alpha} $   with two fitting parameters, $g$ and $\alpha$.  The      exponent is found as $\alpha=0.528 \pm0.013$ and the qubit-cavity coupling as  $g=42 \pm 3$  MHz. 
We note that previously, to our knowledge, the $N^{1/2}$-scaling in the Rabi splitting between two bright states has been observed with ensembles of up to 6 tunable qubits~\cite{Yang-2020}. 
Here, in spite of rising complexity of the measurement setup, we were able to bring into the collective bright states an ensemble of almost 4 times larger number of tunable qubits.

 It should be mentioned that $N^{1/2}$-scaling is robust against errors in the resonance condition, $\nu_c=\epsilon$. The error results in a small ($\sim1/N$) relative deviation of $g_{\rm Rabi}$ from the power law. The estimation follows from  the  expressions for frequencies of bright modes  $\nu_\pm$.

\section{Transmission in a disordered metamaterial}
The most relevant result of this work is measurements of  $S_{21}$ in a mesoscopic metamaterial with large but finite number of qubits $N$ and tunable disorder in their fundamental transition frequencies.
Here, fluctuations are induced by the  diagonal disorder in $\epsilon_j$.  The disorder results in coupling of the pure dark states (with energies close to $\nu_c$) to the cavity mode.  The partial   brightening shifts  randomly the amplitude and phase   of $  S_{21}$.

\begin{figure}[h!]
	\center{\includegraphics[width=\linewidth]{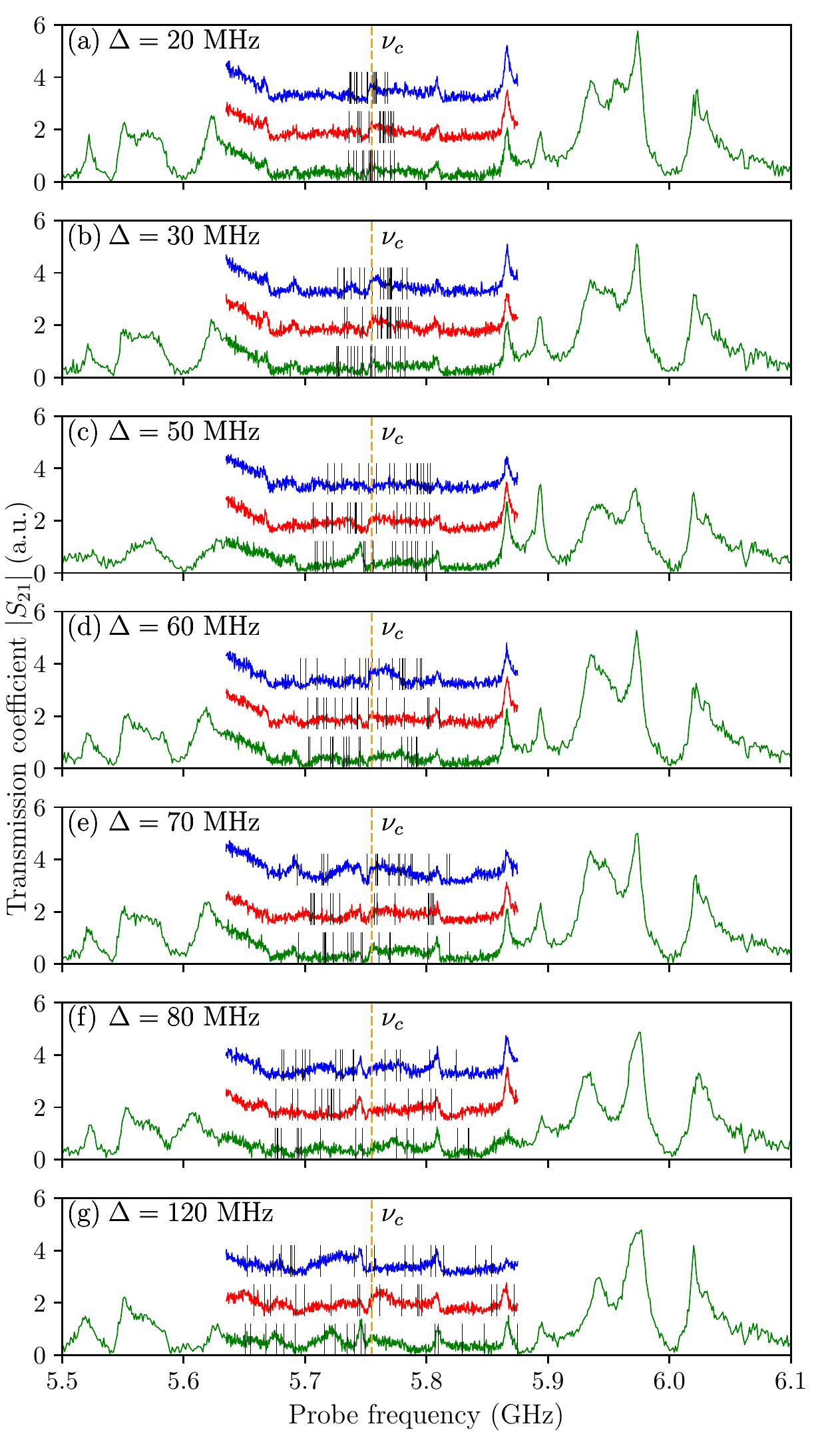}}
	\caption{Microwave   spectroscopy of disordered metamaterial with 17 qubits. The data for transmission coefficient $|S_{21}|$ is presented. Average qubit frequency is tuned into the resonance with the cavity, $\langle\epsilon\rangle=\nu_c$. Panels   (a)-(g) correspond to different spreads $\Delta$ in qubit frequencies.    Different realisations of random frequency offsets are shown in different colors {(green, blue, red)}. Black solid lines denote bare qubit frequencies. Orange dashed lines show the common resonator frequency. }
	\label{fig3}
\end{figure}

Before we analyze fluctuations, we present transmission spectra in the disordered metamaterial. They are shown in   Fig.~\ref{fig3}. 
In these measurements, we used up to 17 qubits which are chosen such that their  individual resonators have  frequencies not very close to  $\nu_c=5.755$~GHz, while the rest of qubits were detuned  down  to 5 GHz and  play no role. 
  As a result, we suppress a coupling to individual resonators in the spectral range of 5.65~--~5.95~GHz. The disorder is introduced artificially by applying random frequency shifts to all qubits with the frequency spreads of $\Delta = 20$, $30$, $50$, $60$, $70$, $80$ and $120$~MHz.  
Transmission data for each probe frequency were averaged over 40 seconds to reduce noise level.
There are peaks at various frequencies in spectra for different disorder realisations. 
According to the above, we are rather certain that we detected dark states manifested by these peaks, and not frequency shifts of individual resonators due to their coupling to the common cavity. Additional measurements  that proof that we detect dark states were also carried out (Appendix~\ref{dark states}).

 Panels (a)-(g) in Fig.~\ref{fig3} correspond to a particular spread of qubit frequencies $\Delta$. There are three curves in each panel (green, red and blue); they correspond  to three particular realizations of disordered $\{\epsilon_j\}_{j=1}^{N}$ (black lines indicate qubit frequencies in each curve). Frequencies $\{\epsilon_j\}_{j=1}^{N}$ are chosen such that they have a particular spread, i.e. $\langle\epsilon_j^2\rangle-\langle\epsilon_j\rangle^2=\Delta^2$. 
Orange dashed line corresponds to the resonator frequency $\nu_{ c}$. Large side peaks on green curves are due to individual qubit readout resonators.

\section{Mesoscopic fluctuations}
\begin{figure}[h!]
	\center{\includegraphics[width=\linewidth]{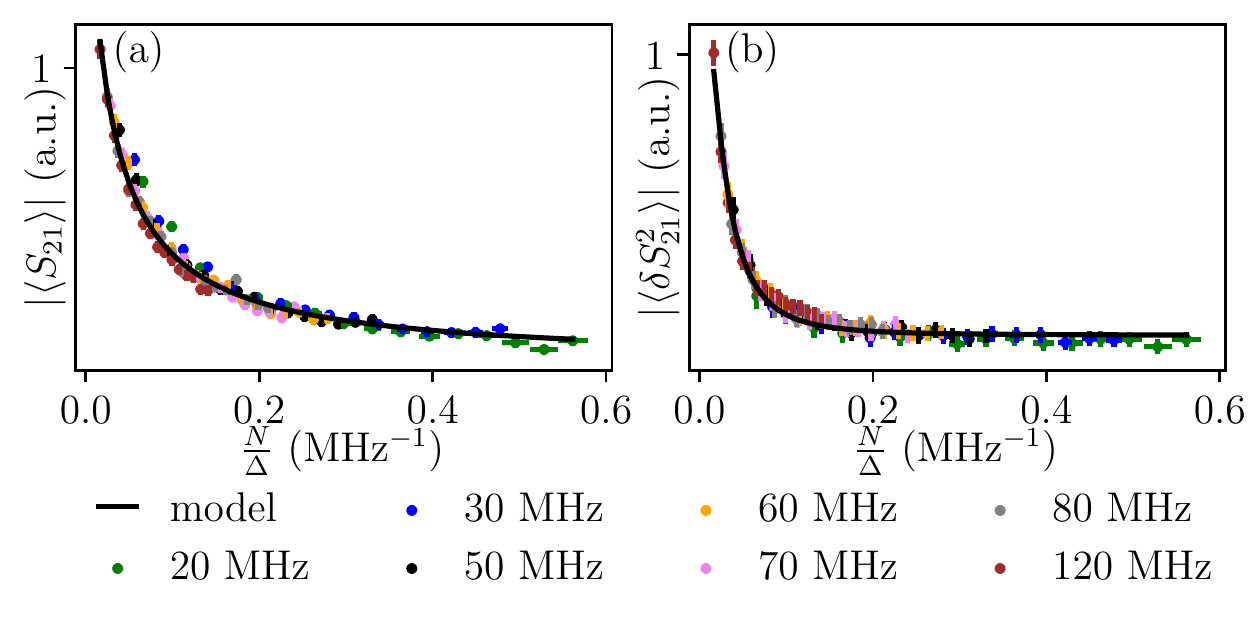}}
	\caption{Results for transmission coefficient and its fluctuations averaged by different  disorder realizations. Different colors stand for  different spreads (see bottom of the figure). (a): averaged transmission amplitude $|\langle S_{21}\rangle|$. (b): averaged  mesoscopic fluctuations  $\langle |\Delta S_{21}|^2 \rangle$. Black lines are fitting theoretical curves given by Eq.~\ref{S12-aver} in (a) and by Eq.~\ref{deltaS12^2} in (b).}
	\label{Results}
\end{figure}

Let us address the average values of $|\langle S_{21}\rangle|$ and $\langle |\Delta S_{21}|^2 \rangle$ in 
the presence of disorder.  Hereafter, the probe signal is   tuned to the cavity mode frequency, $f_p=\nu_c=\langle\epsilon\rangle$, hence,    Eqs.~(\ref{S12-aver}) and (\ref{deltaS12^2}) are applicable. The measurement results for $|\langle S_{21}\rangle|$ and $\langle |\Delta S_{21}|^2 \rangle$ are shown in Figs.~\ref{Results} (a) and (b), respectively. All points are obtained from averaging over 1000 disorder realizations. Different colors correspond to different values of  $\Delta$. 
Here we present the data for $|\langle S_{21}\rangle|$ and $\langle |\Delta S_{21}|^2 \rangle$ as a function of $N/\Delta$ for all spreads and $N$. 
The experimental data is fitted by the formulas $|\langle S_{21}\rangle|$~$=$~$| \frac{a}{(\kappa + \pi g^2  N/\Delta)^\gamma }+ c_1|$ and $\langle |\Delta S_{21}|^2\rangle=\frac{b (N/\Delta)^\beta}{(\kappa + \pi g^2  N/\Delta)^\delta }+c_2$  where the exponents are found as  $\gamma=1.001 \pm 0.005$, $\beta=1.01 \pm 0.02$,  and  $\delta=4 \pm 0.008$.  Their values  show good agreement with theoretical predictions.
Parameters $c_1$ and  $c_2$ are phenomenological corrections that take into account shunting of the circuit probe signal due to interference between the cavity and background transmission (Fano resonance), and thermal noise, which we could not avoid in our measurements. The values of  $\beta$, $\gamma$ and $\delta$ are obtained by least-square fitting of the measurements. We account for the finite values of $c_1$ and $c_2$ using the data processing procedure described in    Appendix~\ref{Processing}. After subtraction of the background scattering ($c_1$ and $c_2$ parameters) we obtain a well agreement
  between the processed data and predicted analytical dependence (black line in Fig.~\ref{Results}).

Eqs.~(\ref{S12-aver}) and (\ref{deltaS12^2}) provide a characteristic number of resonant qubits $N_{0}\sim \frac{\Delta}{2\pi \Gamma}$, above which a  crossover from  mesoscopic behavior   to that of a thermodynamic limit occurs. The estimation follows from the matching condition  $\langle |\Delta S_{21}|^2 \rangle\sim \langle S_{21}  \rangle^2$ where we assume  $\kappa$ is smaller than $g^2 \frac{N}{\Delta}$. (For our setup, $\kappa\approx 30$~MHz and $\Gamma\approx 1$~MHz.) For instance, the measurements with minimal spread $\Delta=20$~MHz provide the estimated value of  $N_{0}\sim 3$--$4$. The data in Fig.~\ref{Results} with  such $N$ and $\Delta$ that have a ratio  $\frac{N}{\Delta}> \frac{1}{2\pi \Gamma}$,  where     $\frac{1}{2\pi \Gamma} 
\approx 
0.16$~MHz$^{-1}$, correspond to values of $N$ that approach the thermodynamic limit formulated above.

\section{Conclusion}
We have studied experimentally an array of 25 tunable transmon qubits coupled to a common resonator. The tunability of qubits allowed us to simulate a diagonal disorder with preassigned distributions. First, we have probed the collective modes of the qubit array by measuring the transmission amplitude of an external microwave signal. By tuning qubits one by one to the resonator frequency, we have observed $N^{1/2}$-scaling law for the Rabi splitting as predicted by the Tavis-Cummings model.   Our most interesting new result is  measurements of the microwave transmission through the qubit metamaterial  in the presence of synthesised disorder in qubit frequencies. We observed mesoscopic fluctuations emerging due to dark states which are very sensitive to disorder  in qubit  frequencies  and their number in the ensemble. 
We observed a decay of the   average value and fluctuations of transmission amplitude  with increasing qubit number  $N$ and decreasing amplitude of the disorder. 
Thus, in the presence of disorder, adding more and more qubits promotes the collective bright state.  The power-law   decay in the transmission    can  evidence for a   semilocalized nature of disordered dark states~\cite{PhysRevB.102.144202} that can be a subject of further investigations.  Our technique thus provides an on-chip quantum simulator of a crossover between the mesoscopic regime and the thermodynamic limit of the Tavis-Cummings model. 

\begin{acknowledgments}
Experimental part of this work was performed with the financial support from the Russian Science Foundation, project N\textsuperscript{\underline{o}}~21-72-30026. D.S.S. and S.V.R.  acknowledge the financial support of the theoretical part of the work by Russian Foundation for Basic Research (RFBR) according to  research project  N\textsuperscript{\underline{o}}~20-37-70028. D.S.S. acknowledges the support  by RFBR  research project  N\textsuperscript{\underline{o}}~20-52-12034, and by DFG Grant No. MI 658/13-1 within a joint DFG-RSF project. W.V.P. acknowledges
a support by RFBR  research project N\textsuperscript{\underline{o}} 19-02-00421.
\end{acknowledgments}

\begin{appendix}

\section{Averaging of the transmission coefficient $S_{21}$}
\label{App:A}
\subsection{Definitions and assumptions}
 \label{Defs}
In this part   we study fluctuations of the transmission coefficient $S_{21}(\omega)$ acquired by the probe signal at the frequency $\omega$. 
 The  transmission coefficient   is a complex valued function   related to    the photon mode Green function   $\mathbf{G}_{\rm ph}(\omega)$ as
\begin{equation}
	S_{21}(\omega)=
	\sqrt{\gamma_{\rm in} \gamma_{\rm out} }\mathbf{G}_{\rm ph}(\omega) \ . \label{S12-Supp}
\end{equation}
Here, loss rates $\gamma_{\rm in}$ and $\gamma_{\rm out}$    determine radiation   from the resonator into {\it in}- and {\it out}-waveguides, respectively. 
Our goal is to calculate fluctuations of    $	S_{21}(\omega)
$ averaged by different disorder realizations in qubit excitation frequencies.

Our calculations are based on  following assumptions:
\begin{enumerate}
	\item 
	The probe signal is small such  the average photon number in the resonator is much smaller than one. This assumption allows to reduce   the Hilbert space  to that of a single excitation (either one photon  or one qubit is excited). Also, this allows to use Tavis-Cummings model in rotating wave approximation.
	
	\item 
	In analytical calculations, we assume that probe frequency $\omega$ is tuned into  a resonance with bare frequency of the resonator mode, {\it i.e.}, $\omega=\nu_c$. This allows us to find compact expressions.
	
	\item 
	We assume   the resonant condition between the resonator mode and all qubits on average as $\nu_c=\epsilon_j$ where $\epsilon_j=\langle \epsilon_j \rangle$ is the averaged by the realizations  frequency of $j$-th qubit ($j\in [1, \ N]$).
	
	\item 
	We assume that disorder distribution functions  are identical for all of the qubits. 
	We suppose that  their dispersions $\Delta_j=\sqrt{\langle \epsilon_j^2 \rangle  -  \langle \epsilon_j \rangle^2}$ are identical, {\it i.e.},   $\Delta_j=\Delta$. 
	
	\item The disorder parameter $\Delta$ is supposed to be known. Also, we know resonator's loss rate, $\kappa$, qubits' loss rates $\Gamma_j=\Gamma$, and identical coupling constants $g_j=g$ between $j$-th qubit and the resonator.
	 
	 \item  The distribution probability $p(\delta\epsilon_j)$ for random qubit detunings $\delta\epsilon_j=\epsilon_j-\langle\epsilon_j\rangle$  is flat 
	 \begin{equation}
	 	p(\delta\epsilon_j)=\frac{1}{\Delta}\theta(\Delta/2-|\delta\epsilon_j|) \ . \label{flat-distr-Supp}
	 \end{equation}
	 Here $\delta \epsilon_j \in [-\Delta/2; \ \Delta/2]$ and $p(\delta\epsilon_j)$ is normalized to unity.
\end{enumerate}

\subsection{ Calculation of the photon's Green function}
\label{GF}
The Tavis-Cummings model (for a particular realization of qubit frequencies $\epsilon_j$ and different couplings $g_j$) reads as 
\begin{equation}
	\hat H =\nu_c \hat a^\dagger \hat a + \sum\limits_{j=1}^N \epsilon_j \hat\sigma_j^+\hat\sigma_j^- + \sum\limits_{j=1}^N g_j( \hat\sigma_j^+ \hat a+ \hat a ^ \dagger \hat\sigma_j^-) \ .
	\label{T-C}
\end{equation}
Here, $\hat a ^\dagger$ and $\hat a$ are the photon creation and annihilation operators,  $ \hat\sigma_j^+$ and $\hat\sigma_j^-$ are raising and lowering operators acting upon $j$-th qubit. 
This Hamiltonian can be represented as $N+1$-dimensional matrix $\mathcal{H}_{i,j}=\langle \psi_i | \hat H |\psi_j\rangle$  after the projection of $\hat H$ on a single excitation basis, $\{|\psi_i\rangle\}_{i=1}^{N+1}= \{ \hat a^\dagger |g.s.\rangle; \ \hat \sigma_1^+ |g.s.\rangle; \ ... ;\  \  \hat\sigma_N^+ |g.s.\rangle \}$:
\begin{equation}
	\mathcal{H}=\begin{bmatrix}
		\nu_c && \mathbf{g}^T \\
		\mathbf g && h_{\rm q}
	\end{bmatrix} \ .
\end{equation}
Here we introduced $N$-dimensional vector $\mathbf {g}=(g_1; \ g_2; \ ...; \ g_N)^T$ and diagonal matrix for qubit ensemble $h_{{\rm q}; i,j}=\delta_{i,j}\epsilon_j$.
The respective Green function is
\begin{equation}
\mathcal{G}_{\rm  }(\omega)=\begin{bmatrix}
G_{\rm ph}^{-1}(\omega) && -\mathbf{g}^T \\
-\mathbf{g} && \mathbf{G}_{{\rm q}}^{-1}(\omega)
\end{bmatrix}^{-1}   \ .  \label{G-Supp}
\end{equation}
It that takes into account loss rates,  
can be written through
the bare Green functions of the lumped resonator and qubit modes, $G_{\rm ph}$ and $G_{{\rm q}}$, respectively.
They read as follows. The resonator' Green function is
\begin{equation}
	G_{\rm ph}(\omega)=\frac{1}{\omega+i\kappa-\nu_c} \ . \end{equation}
Qubits modes are encoded by the diagonal matrix $G_{{\rm q}}$, its elements are $G_{{\rm q}; i,j}=\delta_{i,j} G_{{\rm q}; j}$. They read
\begin{equation}
	G_{{\rm q}; j}(\omega)=\frac{1}{\omega+i\Gamma-\epsilon_j} \ , \ j\in [1, \ N] \ . \end{equation}
The inverse matrix $\mathcal{G}^{-1}$ (\ref{G-Supp}) has non-zero elements on the diagonal, and on the upper row and left column determined by $\mathbf{g}^T$ and $\mathbf{g}$, while other elements are equal to zero. The first diagonal element of $[\mathcal{G}(\omega)]_{1,1}$   corresponds to photon Green function   in the hybrid system, written $\mathcal{G}_{\rm ph}(\omega)$. It is   found  after an expansion by the non-diagonal $\mathbf{g}^T$ and $\mathbf{g}$  in (\ref{G-Supp}) and   following resummation of even order terms. The result is
\begin{equation}
	\mathcal{G}_{\rm ph}(\omega)=\frac{1}{G_{\rm ph}^{-1}(\omega)-\mathbf{g}^T \mathbf{G}_{{\rm q}}(\omega) \mathbf{g}}
\end{equation} 
It takes into account disorders in qubit frequencies and in couplings through the product in the denominator 
\begin{equation}
	\mathbf{g}^T \mathbf{G}_{{\rm q}}(\omega) \mathbf{g}=\sum\limits_{j=1}^N g_j^2G_{{\rm q}; j}(\omega) \ . \end{equation}

\subsection{ Finite size fluctuations of $S_{21}$}
\label{S12-app}
We consider complex valued $S_{21}$ from (\ref{S12-Supp}) at zero detuning ({\textit i.e.} we probe a response at the bare resonator frequency $\nu_c$). It is related to Green function $\mathcal{G}_{\rm ph}(\omega_r) $ and reads:
\begin{multline}
	S_{21}(\omega=\nu_c)= \sqrt{\gamma_{\rm in} \gamma_{\rm out} } \left[g^2 \sum\limits_{j=1}^N\frac{\delta\epsilon_j}{(\delta\epsilon_j)^2 + \Gamma^2} + \right. \\
	+ \left. i\left(\kappa + g^2\Gamma \sum\limits_{j=1}^N\frac{1}{(\delta\epsilon_j)^2 + \Gamma^2}\right)  \right]^{-1} \ . \label{S12-1-Supp}
\end{multline}
Let us analyze the disorder effects starting from a formal limit of infinitely large $N$. Then, the mesoscopic corrections due to $1/N$ with finite $N$ are be calculated. 

There is a self-averaging in the limit of large enough $N$, namely, the sums in (\ref{S12-1-Supp}) are treated as integrals over continuous variable $ \epsilon\in [-\frac{\Delta}{2} \ ; \frac{\Delta}{2}]$ with $p(\epsilon)$ from (\ref{flat-distr-Supp}). This gives:
\begin{equation}
	\sum\limits_{j=1}^N\frac{\delta\epsilon_j}{(\delta\epsilon_j)^2 + \Gamma^2} = N\int p(\epsilon)\frac{ \epsilon d \epsilon}{\epsilon^2 + \Gamma^2}= 0 \ , 
\end{equation}
\begin{equation}
	\sum\limits_{j=1}^N\frac{1}{(\delta\epsilon_j)^2 + \Gamma^2} = N\int p( \epsilon)\frac{d \epsilon}{ \epsilon^2 + \Gamma^2}= \frac{\pi N}{\Delta\Gamma} \ .  \label{Integral}
\end{equation}
(Here, we assumed $\Delta\gg \Gamma$ and calculated the integrals in infinite limits.)
Thus, after this integrations, we find the averaged $\langle S_{21}\rangle$:
\begin{equation}
	\langle S_{21}  \rangle= \sqrt{\gamma_{\rm in} \gamma_{\rm out} } \frac{-i}{\kappa + \pi g^2  N/\Delta  } \ . \label{S12-aver-Supp}
\end{equation}
Now we address the mesoscopic correction to this result due to random variable  given by the first sum, $g^2 \sum\limits_{j=1}^N\frac{\delta\epsilon_j}{(\delta\epsilon_j)^2 + \Gamma^2}$, in the square brackets of (\ref{S12-1-Supp}). We expand $S_{21}$ by the first order in this random valued sum:
\begin{multline}
\left.	S_{21} \approx   \frac{-i \sqrt{\gamma_{\rm in} \gamma_{\rm out} } }{\kappa + \pi g^2  N/\Delta  }\right(1+  \\
+  \frac{i g^2}{\kappa + \pi g^2  N/\Delta}  \sum\limits_{j=1}^N\left.\frac{\delta\epsilon_j}{(\delta\epsilon_j)^2 + \Gamma^2} \right) \ . \label{S12-fluc-Supp}
\end{multline}
According to (\ref{S12-aver-Supp}) and (\ref{S12-fluc-Supp}), we find random deviation $\delta S_{21}$ from averaged value $\langle S_{21}  \rangle$ for a particular realization of values $\epsilon_j$ (we still work with complex valued quantity):
\begin{multline}
	\delta S_{21}=S_{21}- \langle S_{21}  \rangle = \\ 
	=e^{i\varphi_0}   \frac{g^2 \sqrt{\gamma_{\rm in} \gamma_{\rm out} }}{(\kappa + \pi g^2  N/\Delta)^2 }  \sum\limits_{j=1}^N\frac{\delta\epsilon_j}{(\delta\epsilon_j)^2 + \Gamma^2}  \ . \label{deltaS12-Supp}
\end{multline}
At this step we find  its absolute squared and averaged value
\begin{multline}
	\langle |\delta S_{21}|^2 \rangle=  \frac{g^4 \gamma_{\rm in} \gamma_{\rm out}}{(\kappa + \pi g^2  N/\Delta)^4 } \times \\ \times\sum\limits_{i,j=1}^N  \left\langle \frac{\delta\epsilon_i\delta\epsilon_j}{((\delta\epsilon_i)^2 + \Gamma^2)((\delta\epsilon_j)^2 + \Gamma^2)} \right\rangle \ .\label{deltaS12^2-Supp}
\end{multline}
The cross terms with $i\neq j$ in the average (\ref{deltaS12^2-Supp}) cancel out. Hence,
\begin{multline}
  \sum\limits_{i,j=1}^N \left\langle \frac{\delta\epsilon_i\delta\epsilon_j}{((\delta\epsilon_i)^2 + \Gamma^2)((\delta\epsilon_j)^2 + \Gamma^2)} \right\rangle= \\ =
	 \sum\limits_{j=1}^N \left\langle  \frac{\delta\epsilon_i^2}{((\delta\epsilon_j)^2 + \Gamma^2)^2} \right\rangle =  N\int p(\epsilon)\frac{ \epsilon^2 d \epsilon}{(\epsilon^2 + \Gamma^2)^2}= \\ =\frac{\pi N}{2\Gamma \Delta} \ . \label{aver-2-Supp}
\end{multline}
(We note, that $\langle |\delta S_{21}|^2 \rangle$ decays  faster than $\langle S_{21}\rangle$ at large $N$ that means the expansion in (\ref{S12-fluc-Supp}) is a controllable approximation.)
Finally, we find {\it relative} mesoscopic fluctuations of $S_{21}$, combining (\ref{S12-aver-Supp}), (\ref{deltaS12^2-Supp}) and (\ref{aver-2-Supp}), which reads:
\begin{equation}
	\frac{\sqrt{\langle |\delta S_{21}|^2 \rangle}}{|\langle  S_{21}\rangle|}=  \frac{g^2}{\kappa + \pi g^2  N/\Delta }   \sqrt{\frac{\pi N}{2\Gamma \Delta}}  \ . \label{S12-fluc-main-Supp}
\end{equation}
We note that if one assumes that the resonator's relaxation is small, $\kappa\ll  g^2  N/\Delta $,  we find the following scaling where $g$ does not appear:
\begin{equation}
	\frac{\sqrt{\langle |\delta S_{21}|^2 \rangle}}{|\langle  S_{21}\rangle|}=   \sqrt{\frac{1}{ N} \frac{\Delta}{2\pi \Gamma }}  \ . \label{deltaS12-fluct-main-1-Supp}
\end{equation}
However, experimental observation of this scaling requires large $N$ and small $\kappa$ which are not realized in our device.

\section{Sample fabrication}\label{Sample fabrication}
\begin{figure}[h!]
	\center{\includegraphics[width=\linewidth]{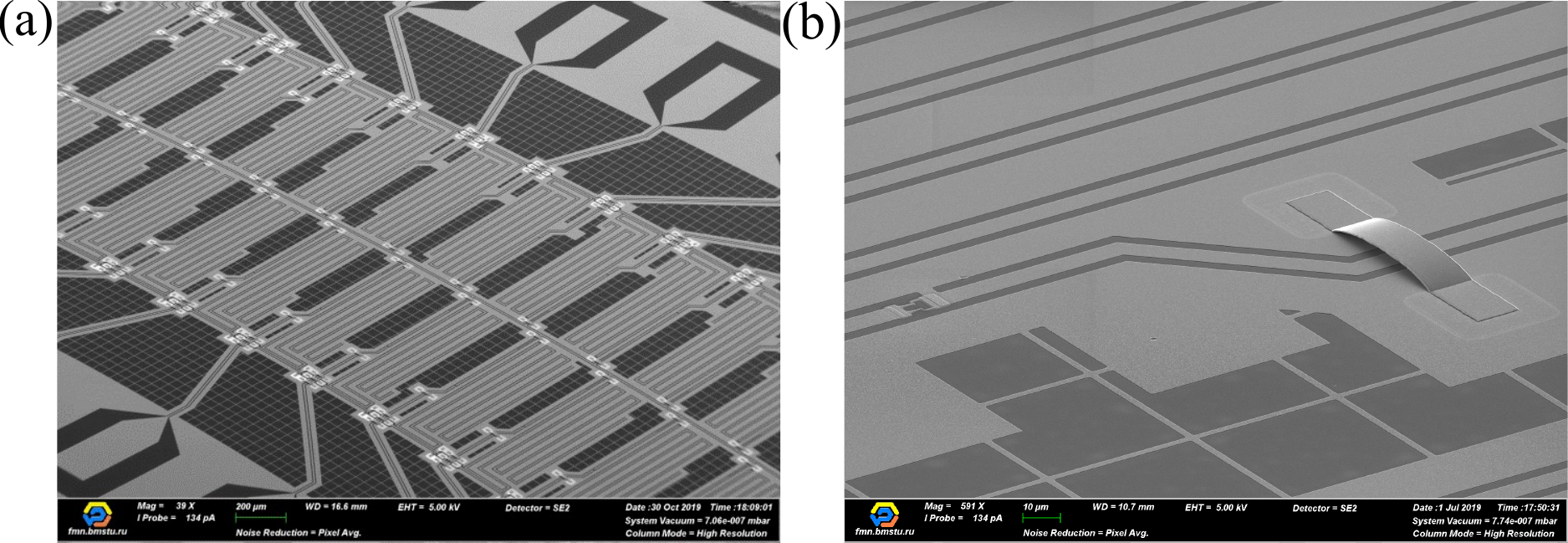}}
	\caption{(a): SEM image of a twenty five qubits array (b): SEM image of a Josephson junction SQUID with low impedance crossover.}
	\label{figSEM}
\end{figure}

To fabricate the superconducting chip based on the array of 25 transmon qubits coupled to a common coplanar resonator we use the three-stage process.
The process includes following stages: I) epitaxial Al base layer deposition using two-step SCULL process \cite{Rodionov2019ER}
and Al patterning with laser direct lithography (including transmon capacitor ground plane, waveguides, resonators and flux bias lines); II) double-angle evaporation of Josephson junctions followed by lift-off; 
III) resist-based low impedance crossover fabrication.
SEM images of metamaterial and enlarged fragment of a Josephson junction SQUID are shown in Fig. \ref{figSEM}.

The fabrication process starts with multi-step wet chemical cleaning of a high-resistivity intrinsic silicon sample ($\rho > 10000$ $\Omega\cdot$cm, 525 $\mu$m thick) in a Piranha solution (1:4) followed by native oxide removal in HF (1:50) for 120 seconds.
Immediately after a 100 nm thick epitaxial Al base layer is deposited with UHV e-beam evaporation SCULL technique \cite{Rodionov2019ER} followed by its direct laser lithography patterning and dry etching in BCl$_3$/Cl$_2$-based gasses.
Than a two-layer e-beam resists stack (300 nm thick PMMA e-beam resist on top of a 500 nm thick MMA copolymer) is spin coated followed by 50 kV e-beam exposure.
After development and oxygen plasma treatment, we performed UHV e-beam shadow evaporation of Al-AlOx-Al Josephson junctions (62$^{\rm o}$/0$^{\rm o}$, 25/45 nm). Low impedance free-standing crossovers are fabricated by means of a four-step \cite{doi:10.1063/1.4863745}
process: (I) crossovers pads laser lithography, (II) 300 nm thick Al film e-beam deposition, (III) crossovers topology laser lithography; (IV) BCl$_3$/Cl$_2$-based dry plasma etching. Finally, we stripped both resist layers in an NMP-based solvent.

\begin{figure*}[t!]
	\center{\includegraphics[width=\linewidth]{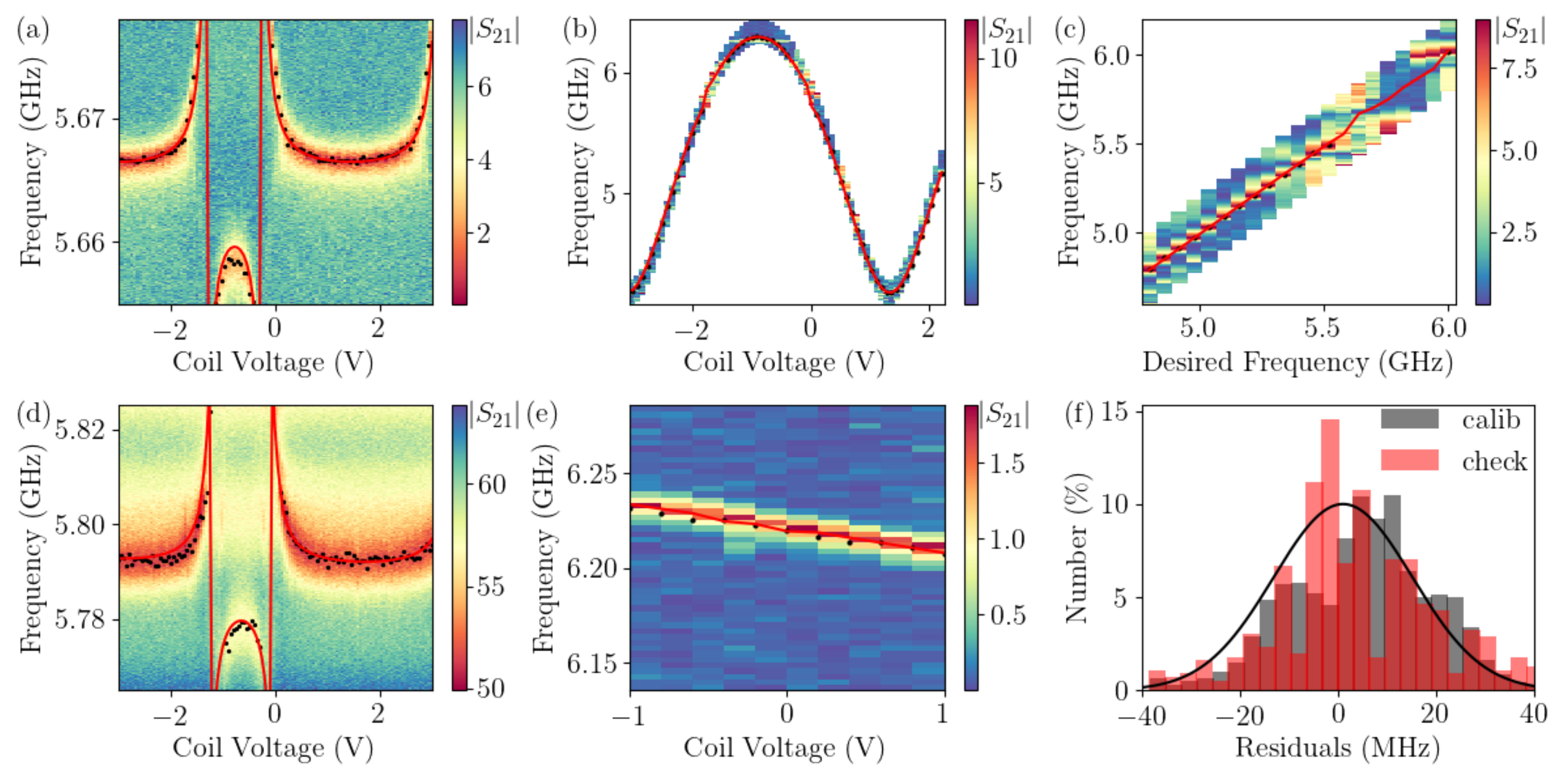}}
	\caption{Typical spectra used to calibration and fitting residuals. (a): Single-tone spectroscopy of individual resonator. (b): Adaptive two-tone spectroscopy. (c): Check accuracy of setting qubits into desired frequency. (d): Single-tone spectroscopy of common cavity. (e): Two-tone spectroscopy tuning voltage of coil connected to another qubit. (f): Distributions of residuals from two-tone spectra used for calibration or check frequencies. Solid black line was calculated from dispersion and average of distribution. (a)-(e) Red solid lines were calculated using Tavis-Cummings model, black dots is data used for fitting task.}
	\label{figcal1}
\end{figure*}

\section{ Calibration}\label{Calibration}

Spectroscopic measurements have been performed to determine device parameters such as Josephson energies of SQUID junctions, capacitive energies, coupling between qubits and common cavity, qubits and individual resonators, mutual inductance between coils and qubits, frequencies of resonators, flux biases of the SQUIDs.
Typical results of these measurements are presented in Fig. \ref{figcal1} (a), (b), (d), (e). \par
For each flux configuration, the frequency corresponding to maximal two-tone response is taken as qubit frequency.
Outlier points are dropped after we extract device parameters by fitting data with the Tavis-Cummings model (\ref{T-C}) extended with the individual resonators ($\nu_j$). We also consider the dependence of the coupling coefficients to common cavity ($g_j=k_j\sqrt{\epsilon_j}$) and to individual readout resonators ($g_j^{ind}=k_j^{ind}\sqrt{\epsilon_j}$) on qubit frequencies ($\epsilon_j$): 
\begin{multline}
\hat{H}= \sum_{j=1}^{N}\nu_{j}\hat{a}^{\dag}_{j}\hat{a}_{j}+\sum_{j=1}^{N}\epsilon_j\hat{\sigma}_j^+\hat{\sigma}_j^-+\nu_{c}\hat{a}^{\dag}\hat{a} + \\ +\sum_{j=1}^{N}k_{j}^{ind}\sqrt{\epsilon_j}(\hat{a}^{\dag}_{j}+\hat{a}_{j})(\hat{\sigma}^{+}_j+\hat{\sigma}^{-}_j)+\\
 +\sum_{j=1}^{N}k_j\sqrt{\epsilon_j}(\hat{a}^{\dag}+\hat{a})(\hat{\sigma}^{+}_j+\hat{\sigma}^{-}_j) \ , 
\label{T-C}
\end{multline}
For the dependence of bare transmon frequency on SQUID flux we use the following formula:
\begin{multline}
\epsilon(\phi)=\sqrt{8E_C}((E_{J1}+E_{J2})^2\cos^2\phi+ \\ +(E_{J1}-E_{J2})^2\sin^2\phi)^{\frac{1}{4}}  - E_C,  
\label{fqb}
\end{multline}
where $E_{J1}$ and $E_{J2}$ are Josephson energies of   DC SQUIDs' junctions, $E_C$ is the transmon charging energy, $\phi=2\pi\frac{\Phi}{\Phi_0}$ is the dimensionless magnetic flux threaded by the SQUID.
We assume linear dependence of SQUID fluxes on DC voltage applied to the coils,
\begin{equation}
\phi_i=\sum L_{ij}V_j +\phi_i^0 \ ,
\label{flux}
\end{equation}
where index $i$ corresponds to qubit number, $j$ to coil number, $\phi^0$ is frozen dimensionless magnetic flux in the SQUID.  
The effective eigenmode of a specific transmon was chosen as the frequency of the mode with the largest participation in this transmon.
The fitting was performed with a least-squares cost function. The  standard deviation is 20 MHz.  
The distribution of residuals is shown on Fig.~\ref{figcal1}~(f) (black bar).

When setting a transmon to some frequency, we set not the bare frequency of the transmon, but the eigenfrequency of the transmon-individual resonator system. 
The bare qubit frequency is calculated using a coupled linear oscillator model $\epsilon_c=\frac{\epsilon+\nu}{2}\pm\frac{\sqrt{(\epsilon-\nu)^2+4k\epsilon}}{2}$, where $\epsilon_c$ corresponds to the eigenfrequency of the coupled system.
After that, fluxes are calculated from equation (\ref{fqb}) and, then, the linear system of equations (\ref{flux}) is solved.

To evaluate frequency control accuracy independently we perform two-tone spectroscopy measurement setting all qubits to equal desired frequency and then tuning one of them in some range, while others are persisted in their position.
A typical result of these measurements and the distribution of residuals (red bar) are shown on Fig. \ref{figcal1} (c) and (f), correspondingly.
The standard deviation is also equal to 20~MHz. We consider this value as the frequency control error.

 \section{ Detection of dark states}\label{dark states}
\begin{figure}[h!]
	\center{\includegraphics[width=9 cm]{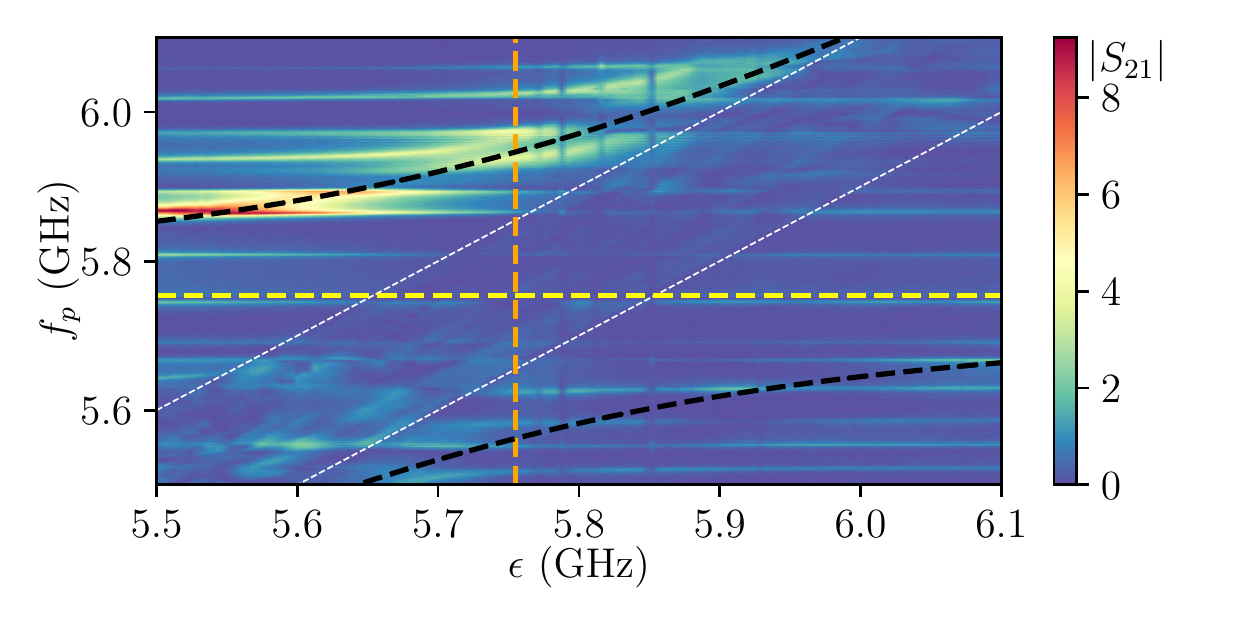}}
	\caption{ Transmission spectrum for detection of dark states. Yellow dashed line stands for bare cavity mode frequency $\nu_c$. Black dashed lines are bare frequencies of tunable qubits. Orange dashed line corresponds to the measurements presented in Fig.~\ref{fig3}, when the central frequency of the qubit ensemble is equal to the bare frequency of the common cavity. White dashed lines show the area of possible qubits frequencies.} 
	\label{detect}
\end{figure}

We carried out measurements to ensure that dark states are observable in our experiments.
We tuned the central frequency of the ensemble of qubits with fixed disorder and provide transmission measurements.
The results of these measurements are presented in Fig. \ref{detect}.
White dashed lines show the bounds for the qubit frequencies in the ensemble.
Several continuous lines are visible inside white bounds that are parallel to the white lines.
This fact reveals that the system response is changed as the central frequency of the qubit ensemble, i.e., each line corresponds to a certain dark state.
This is an argument to successful detection of dark states for data presented in Fig.~\ref{fig3}.
Also, Fig. \ref{detect} shows that individual resonator frequencies inside the range between 5.65 GHz and 5.95 GHz do not depend on the central frequency of the qubit ensemble, as pointed out in the main text.

\begin{figure*} [t!]
	\center{\includegraphics[width=\linewidth]{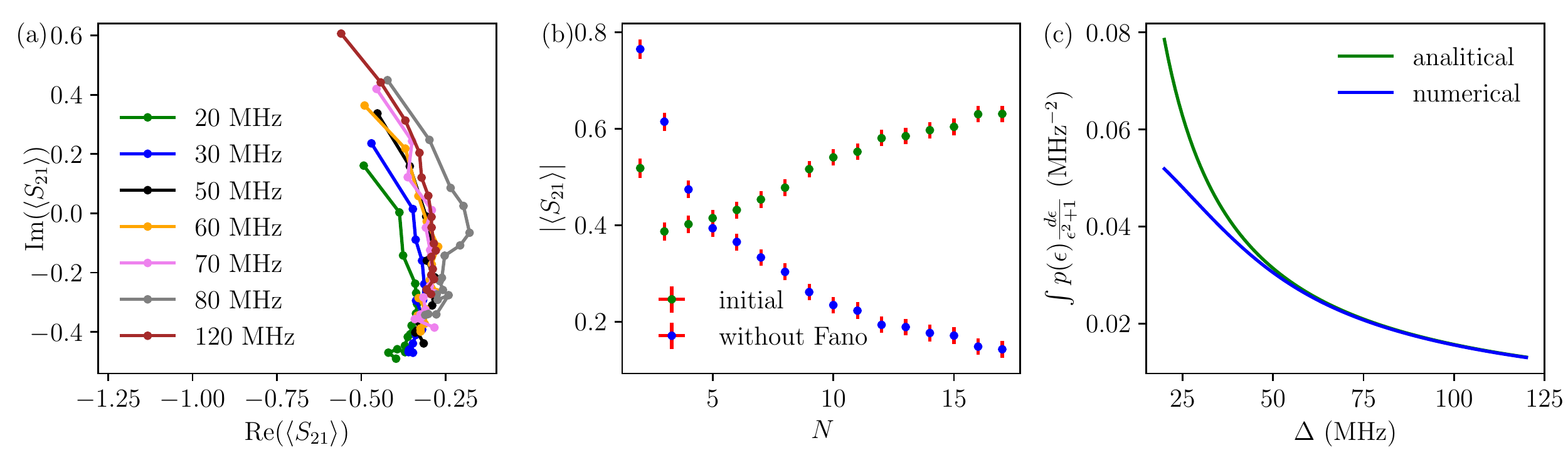}}
	\caption{Processing of experimental data. (a) - average of transmission amplitude on complex plane from the number of resonant qubits for different distribution spreads $\Delta$. (b) - dependence of the absolute value of average transmission amplitude on $N$ for initial experimental data and after elimination the Fano resonance, (c) -  the integral (\ref{Integral}) value dependence on the random distribution spread calculated analytically and numerically to account for calibration errors.}
	\label{Proc}
\end{figure*}

\section{ Processing of experimental data}
\label{Processing}
 The variance in experimental data can be due to various reasons as, e.g., specific realizations of flat distribution, calibration of qubit frequencies, interference  between the cavity and background transmission (Fano resonance), and thermal noise. The influence of some of these factors is discussed below. 

Average transmission amplitudes for different spreads $\Delta$ are shown in Fig. \ref{Proc} (a). From~\eqref{S12-aver-Supp} we expect a straight line dependence for real and imaginary parts of $S_{21}$ with increasing number of qubits $N$ for each value of $\Delta$.  
It is seen from the Fig.~\ref{Proc} (a), that experimental curves are close to straight lines but tend to the same nonzero value. 
As it mentioned in the main text one reason of such behaviour is the appearance of Fano resonance. In our post processing we had to introduce phenomenological corrections to eliminate this parasitic effect. Corrections are selected the same for all curves.
A typical example of such processing is shown in Fig.~\ref{Proc} (b), where you can see the dependence  of the absolute value of the transmission signal on the number of resonant qubits for the initial data and for the data after elimination the Fano resonance influence.

An additional processing step was made to take into account the difference between the real distribution and the random flat distribution due to calibration errors Fig. \ref{figcal1} (f).
These errors prevent us from using a simple analytical calculation of the integral in~(\ref{Integral}).
The comparison of the analytical and numerical solution of this integral for different  spreads $\Delta$ is shown in Fig.~\ref{Proc}~(c).
A significant discrepancy is observed at small values of $\Delta$. For subsequent calculations presented in the main text we used a numerical approach, in order to determine the effective width of distribution.

\begin{figure} [h!]
	\center{\includegraphics[width=0.95\linewidth]{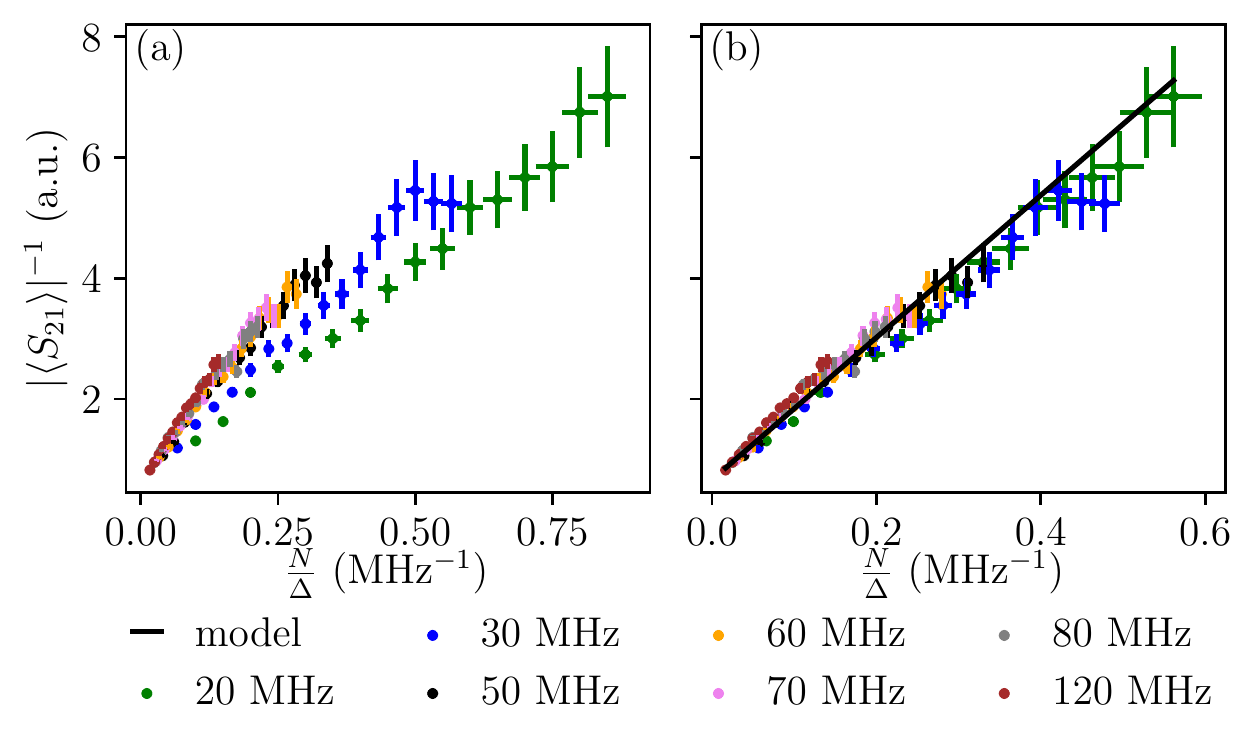}}
	\caption{Inverted transmission coefficient $|1/\langle S_{21}\rangle|$ as a function of the ratio $N/\Delta$. Different colors stand for  different spreads (see bottom of the figure). (a) - the raw  data with initial $\Delta$, (b) - processed data with corrected $\Delta$. Black line  in (b) is the fitting  curve which is given by Eq.~\ref{S12-aver-Supp}.}
	\label{Delts}
\end{figure}

The influence of these errors is shown in Fig.~\ref{Delts}, where presented inverted averaged transmission coefficient $|1/\langle S_{21}\rangle|$ for initial $\Delta$ (a) and corrected (b).

According to  Eq.~(\ref{S12-aver-Supp}), there should be a linear dependence between the inverted transmission $|1/\langle S_{21}\rangle|$ and the ratio $N/\Delta$, up to a constant shift.
The raw data is well fitted by different lines for each width of distribution, while processed data - by one line.
\end{appendix}
 

\begin{thebibliography}{39}%
\makeatletter
\providecommand \@ifxundefined [1]{%
 \@ifx{#1\undefined}
}%
\providecommand \@ifnum [1]{%
 \ifnum #1\expandafter \@firstoftwo
 \else \expandafter \@secondoftwo
 \fi
}%
\providecommand \@ifx [1]{%
 \ifx #1\expandafter \@firstoftwo
 \else \expandafter \@secondoftwo
 \fi
}%
\providecommand \natexlab [1]{#1}%
\providecommand \enquote  [1]{``#1''}%
\providecommand \bibnamefont  [1]{#1}%
\providecommand \bibfnamefont [1]{#1}%
\providecommand \citenamefont [1]{#1}%
\providecommand \href@noop [0]{\@secondoftwo}%
\providecommand \href [0]{\begingroup \@sanitize@url \@href}%
\providecommand \@href[1]{\@@startlink{#1}\@@href}%
\providecommand \@@href[1]{\endgroup#1\@@endlink}%
\providecommand \@sanitize@url [0]{\catcode `\\12\catcode `\$12\catcode
  `\&12\catcode `\#12\catcode `\^12\catcode `\_12\catcode `\%12\relax}%
\providecommand \@@startlink[1]{}%
\providecommand \@@endlink[0]{}%
\providecommand \url  [0]{\begingroup\@sanitize@url \@url }%
\providecommand \@url [1]{\endgroup\@href {#1}{\urlprefix }}%
\providecommand \urlprefix  [0]{URL }%
\providecommand \Eprint [0]{\href }%
\providecommand \doibase [0]{http://dx.doi.org/}%
\providecommand \selectlanguage [0]{\@gobble}%
\providecommand \bibinfo  [0]{\@secondoftwo}%
\providecommand \bibfield  [0]{\@secondoftwo}%
\providecommand \translation [1]{[#1]}%
\providecommand \BibitemOpen [0]{}%
\providecommand \bibitemStop [0]{}%
\providecommand \bibitemNoStop [0]{.\EOS\space}%
\providecommand \EOS [0]{\spacefactor3000\relax}%
\providecommand \BibitemShut  [1]{\csname bibitem#1\endcsname}%
\let\auto@bib@innerbib\@empty
\bibitem [{\citenamefont {Arute}\ \emph {et~al.}(2019)\citenamefont {Arute},
  \citenamefont {Arya}, \citenamefont {Babbush}, \citenamefont {Bacon},
  \citenamefont {Bardin}, \citenamefont {Barends}, \citenamefont {Biswas},
  \citenamefont {Boixo}, \citenamefont {Brandao}, \citenamefont {Buell},
  \citenamefont {Burkett}, \citenamefont {Chen}, \citenamefont {Chen},
  \citenamefont {Chiaro}, \citenamefont {Collins}, \citenamefont {Courtney},
  \citenamefont {Dunsworth}, \citenamefont {Farhi}, \citenamefont {Foxen},
  \citenamefont {Fowler}, \citenamefont {Gidney}, \citenamefont {Giustina},
  \citenamefont {Graff}, \citenamefont {Guerin}, \citenamefont {Habegger},
  \citenamefont {Harrigan}, \citenamefont {Hartmann}, \citenamefont {Ho},
  \citenamefont {Hoffmann}, \citenamefont {Huang}, \citenamefont {Humble},
  \citenamefont {Isakov}, \citenamefont {Jeffrey}, \citenamefont {Jiang},
  \citenamefont {Kafri}, \citenamefont {Kechedzhi}, \citenamefont {Kelly},
  \citenamefont {Klimov}, \citenamefont {Knysh}, \citenamefont {Korotkov},
  \citenamefont {Kostritsa}, \citenamefont {Landhuis}, \citenamefont
  {Lindmark}, \citenamefont {Lucero}, \citenamefont {Lyakh}, \citenamefont
  {Mandr{\`a}}, \citenamefont {McClean}, \citenamefont {McEwen}, \citenamefont
  {Megrant}, \citenamefont {Mi}, \citenamefont {Michielsen}, \citenamefont
  {Mohseni}, \citenamefont {Mutus}, \citenamefont {Naaman}, \citenamefont
  {Neeley}, \citenamefont {Neill}, \citenamefont {Niu}, \citenamefont {Ostby},
  \citenamefont {Petukhov}, \citenamefont {Platt}, \citenamefont {Quintana},
  \citenamefont {Rieffel}, \citenamefont {Roushan}, \citenamefont {Rubin},
  \citenamefont {Sank}, \citenamefont {Satzinger}, \citenamefont {Smelyanskiy},
  \citenamefont {Sung}, \citenamefont {Trevithick}, \citenamefont
  {Vainsencher}, \citenamefont {Villalonga}, \citenamefont {White},
  \citenamefont {Yao}, \citenamefont {Yeh}, \citenamefont {Zalcman},
  \citenamefont {Neven},\ and\ \citenamefont {Martinis}}]{Arute:2019aa}%
  \BibitemOpen
  \bibfield  {author} {\bibinfo {author} {\bibfnamefont {F.}~\bibnamefont
  {Arute}}, \bibinfo {author} {\bibfnamefont {K.}~\bibnamefont {Arya}},
  \bibinfo {author} {\bibfnamefont {R.}~\bibnamefont {Babbush}}, \bibinfo
  {author} {\bibfnamefont {D.}~\bibnamefont {Bacon}}, \bibinfo {author}
  {\bibfnamefont {J.~C.}\ \bibnamefont {Bardin}}, \bibinfo {author}
  {\bibfnamefont {R.}~\bibnamefont {Barends}}, \bibinfo {author} {\bibfnamefont
  {R.}~\bibnamefont {Biswas}}, \bibinfo {author} {\bibfnamefont
  {S.}~\bibnamefont {Boixo}}, \bibinfo {author} {\bibfnamefont {F.~G. S.~L.}\
  \bibnamefont {Brandao}}, \bibinfo {author} {\bibfnamefont {D.~A.}\
  \bibnamefont {Buell}}, \bibinfo {author} {\bibfnamefont {B.}~\bibnamefont
  {Burkett}}, \bibinfo {author} {\bibfnamefont {Y.}~\bibnamefont {Chen}},
  \bibinfo {author} {\bibfnamefont {Z.}~\bibnamefont {Chen}}, \bibinfo {author}
  {\bibfnamefont {B.}~\bibnamefont {Chiaro}}, \bibinfo {author} {\bibfnamefont
  {R.}~\bibnamefont {Collins}}, \bibinfo {author} {\bibfnamefont
  {W.}~\bibnamefont {Courtney}}, \bibinfo {author} {\bibfnamefont
  {A.}~\bibnamefont {Dunsworth}}, \bibinfo {author} {\bibfnamefont
  {E.}~\bibnamefont {Farhi}}, \bibinfo {author} {\bibfnamefont
  {B.}~\bibnamefont {Foxen}}, \bibinfo {author} {\bibfnamefont
  {A.}~\bibnamefont {Fowler}}, \bibinfo {author} {\bibfnamefont
  {C.}~\bibnamefont {Gidney}}, \bibinfo {author} {\bibfnamefont
  {M.}~\bibnamefont {Giustina}}, \bibinfo {author} {\bibfnamefont
  {R.}~\bibnamefont {Graff}}, \bibinfo {author} {\bibfnamefont
  {K.}~\bibnamefont {Guerin}}, \bibinfo {author} {\bibfnamefont
  {S.}~\bibnamefont {Habegger}}, \bibinfo {author} {\bibfnamefont {M.~P.}\
  \bibnamefont {Harrigan}}, \bibinfo {author} {\bibfnamefont {M.~J.}\
  \bibnamefont {Hartmann}}, \bibinfo {author} {\bibfnamefont {A.}~\bibnamefont
  {Ho}}, \bibinfo {author} {\bibfnamefont {M.}~\bibnamefont {Hoffmann}},
  \bibinfo {author} {\bibfnamefont {T.}~\bibnamefont {Huang}}, \bibinfo
  {author} {\bibfnamefont {T.~S.}\ \bibnamefont {Humble}}, \bibinfo {author}
  {\bibfnamefont {S.~V.}\ \bibnamefont {Isakov}}, \bibinfo {author}
  {\bibfnamefont {E.}~\bibnamefont {Jeffrey}}, \bibinfo {author} {\bibfnamefont
  {Z.}~\bibnamefont {Jiang}}, \bibinfo {author} {\bibfnamefont
  {D.}~\bibnamefont {Kafri}}, \bibinfo {author} {\bibfnamefont
  {K.}~\bibnamefont {Kechedzhi}}, \bibinfo {author} {\bibfnamefont
  {J.}~\bibnamefont {Kelly}}, \bibinfo {author} {\bibfnamefont {P.~V.}\
  \bibnamefont {Klimov}}, \bibinfo {author} {\bibfnamefont {S.}~\bibnamefont
  {Knysh}}, \bibinfo {author} {\bibfnamefont {A.}~\bibnamefont {Korotkov}},
  \bibinfo {author} {\bibfnamefont {F.}~\bibnamefont {Kostritsa}}, \bibinfo
  {author} {\bibfnamefont {D.}~\bibnamefont {Landhuis}}, \bibinfo {author}
  {\bibfnamefont {M.}~\bibnamefont {Lindmark}}, \bibinfo {author}
  {\bibfnamefont {E.}~\bibnamefont {Lucero}}, \bibinfo {author} {\bibfnamefont
  {D.}~\bibnamefont {Lyakh}}, \bibinfo {author} {\bibfnamefont
  {S.}~\bibnamefont {Mandr{\`a}}}, \bibinfo {author} {\bibfnamefont {J.~R.}\
  \bibnamefont {McClean}}, \bibinfo {author} {\bibfnamefont {M.}~\bibnamefont
  {McEwen}}, \bibinfo {author} {\bibfnamefont {A.}~\bibnamefont {Megrant}},
  \bibinfo {author} {\bibfnamefont {X.}~\bibnamefont {Mi}}, \bibinfo {author}
  {\bibfnamefont {K.}~\bibnamefont {Michielsen}}, \bibinfo {author}
  {\bibfnamefont {M.}~\bibnamefont {Mohseni}}, \bibinfo {author} {\bibfnamefont
  {J.}~\bibnamefont {Mutus}}, \bibinfo {author} {\bibfnamefont
  {O.}~\bibnamefont {Naaman}}, \bibinfo {author} {\bibfnamefont
  {M.}~\bibnamefont {Neeley}}, \bibinfo {author} {\bibfnamefont
  {C.}~\bibnamefont {Neill}}, \bibinfo {author} {\bibfnamefont {M.~Y.}\
  \bibnamefont {Niu}}, \bibinfo {author} {\bibfnamefont {E.}~\bibnamefont
  {Ostby}}, \bibinfo {author} {\bibfnamefont {A.}~\bibnamefont {Petukhov}},
  \bibinfo {author} {\bibfnamefont {J.~C.}\ \bibnamefont {Platt}}, \bibinfo
  {author} {\bibfnamefont {C.}~\bibnamefont {Quintana}}, \bibinfo {author}
  {\bibfnamefont {E.~G.}\ \bibnamefont {Rieffel}}, \bibinfo {author}
  {\bibfnamefont {P.}~\bibnamefont {Roushan}}, \bibinfo {author} {\bibfnamefont
  {N.~C.}\ \bibnamefont {Rubin}}, \bibinfo {author} {\bibfnamefont
  {D.}~\bibnamefont {Sank}}, \bibinfo {author} {\bibfnamefont {K.~J.}\
  \bibnamefont {Satzinger}}, \bibinfo {author} {\bibfnamefont {V.}~\bibnamefont
  {Smelyanskiy}}, \bibinfo {author} {\bibfnamefont {K.~J.}\ \bibnamefont
  {Sung}}, \bibinfo {author} {\bibfnamefont {M.~D.}\ \bibnamefont
  {Trevithick}}, \bibinfo {author} {\bibfnamefont {A.}~\bibnamefont
  {Vainsencher}}, \bibinfo {author} {\bibfnamefont {B.}~\bibnamefont
  {Villalonga}}, \bibinfo {author} {\bibfnamefont {T.}~\bibnamefont {White}},
  \bibinfo {author} {\bibfnamefont {Z.~J.}\ \bibnamefont {Yao}}, \bibinfo
  {author} {\bibfnamefont {P.}~\bibnamefont {Yeh}}, \bibinfo {author}
  {\bibfnamefont {A.}~\bibnamefont {Zalcman}}, \bibinfo {author} {\bibfnamefont
  {H.}~\bibnamefont {Neven}}, \ and\ \bibinfo {author} {\bibfnamefont {J.~M.}\
  \bibnamefont {Martinis}},\ }\href {\doibase 10.1038/s41586-019-1666-5}
  {\bibfield  {journal} {\bibinfo  {journal} {Nature}\ }\textbf {\bibinfo
  {volume} {574}},\ \bibinfo {pages} {505} (\bibinfo {year}
  {2019})}\BibitemShut {NoStop}%
\bibitem [{\citenamefont {Clerk}\ \emph {et~al.}(2020)\citenamefont {Clerk},
  \citenamefont {Lehnert}, \citenamefont {Bertet}, \citenamefont {Petta},\ and\
  \citenamefont {Nakamura}}]{Clerk:2020aa}%
  \BibitemOpen
  \bibfield  {author} {\bibinfo {author} {\bibfnamefont {A.~A.}\ \bibnamefont
  {Clerk}}, \bibinfo {author} {\bibfnamefont {K.~W.}\ \bibnamefont {Lehnert}},
  \bibinfo {author} {\bibfnamefont {P.}~\bibnamefont {Bertet}}, \bibinfo
  {author} {\bibfnamefont {J.~R.}\ \bibnamefont {Petta}}, \ and\ \bibinfo
  {author} {\bibfnamefont {Y.}~\bibnamefont {Nakamura}},\ }\href {\doibase
  10.1038/s41567-020-0797-9} {\bibfield  {journal} {\bibinfo  {journal} {Nature
  Physics}\ }\textbf {\bibinfo {volume} {16}},\ \bibinfo {pages} {257}
  (\bibinfo {year} {2020})}\BibitemShut {NoStop}%
\bibitem [{\citenamefont {Barends}\ \emph {et~al.}(2015)\citenamefont
  {Barends}, \citenamefont {Lamata}, \citenamefont {Kelly}, \citenamefont
  {Garc{\'\i}a-{\'A}lvarez}, \citenamefont {Fowler}, \citenamefont {Megrant},
  \citenamefont {Jeffrey}, \citenamefont {White}, \citenamefont {Sank},
  \citenamefont {Mutus}, \citenamefont {Campbell}, \citenamefont {Chen},
  \citenamefont {Chen}, \citenamefont {Chiaro}, \citenamefont {Dunsworth},
  \citenamefont {Hoi}, \citenamefont {Neill}, \citenamefont {O'Malley},
  \citenamefont {Quintana}, \citenamefont {Roushan}, \citenamefont
  {Vainsencher}, \citenamefont {Wenner}, \citenamefont {Solano},\ and\
  \citenamefont {Martinis}}]{Barends:2015aa}%
  \BibitemOpen
  \bibfield  {author} {\bibinfo {author} {\bibfnamefont {R.}~\bibnamefont
  {Barends}}, \bibinfo {author} {\bibfnamefont {L.}~\bibnamefont {Lamata}},
  \bibinfo {author} {\bibfnamefont {J.}~\bibnamefont {Kelly}}, \bibinfo
  {author} {\bibfnamefont {L.}~\bibnamefont {Garc{\'\i}a-{\'A}lvarez}},
  \bibinfo {author} {\bibfnamefont {A.~G.}\ \bibnamefont {Fowler}}, \bibinfo
  {author} {\bibfnamefont {A.}~\bibnamefont {Megrant}}, \bibinfo {author}
  {\bibfnamefont {E.}~\bibnamefont {Jeffrey}}, \bibinfo {author} {\bibfnamefont
  {T.~C.}\ \bibnamefont {White}}, \bibinfo {author} {\bibfnamefont
  {D.}~\bibnamefont {Sank}}, \bibinfo {author} {\bibfnamefont {J.~Y.}\
  \bibnamefont {Mutus}}, \bibinfo {author} {\bibfnamefont {B.}~\bibnamefont
  {Campbell}}, \bibinfo {author} {\bibfnamefont {Y.}~\bibnamefont {Chen}},
  \bibinfo {author} {\bibfnamefont {Z.}~\bibnamefont {Chen}}, \bibinfo {author}
  {\bibfnamefont {B.}~\bibnamefont {Chiaro}}, \bibinfo {author} {\bibfnamefont
  {A.}~\bibnamefont {Dunsworth}}, \bibinfo {author} {\bibfnamefont {I.~C.}\
  \bibnamefont {Hoi}}, \bibinfo {author} {\bibfnamefont {C.}~\bibnamefont
  {Neill}}, \bibinfo {author} {\bibfnamefont {P.~J.~J.}\ \bibnamefont
  {O'Malley}}, \bibinfo {author} {\bibfnamefont {C.}~\bibnamefont {Quintana}},
  \bibinfo {author} {\bibfnamefont {P.}~\bibnamefont {Roushan}}, \bibinfo
  {author} {\bibfnamefont {A.}~\bibnamefont {Vainsencher}}, \bibinfo {author}
  {\bibfnamefont {J.}~\bibnamefont {Wenner}}, \bibinfo {author} {\bibfnamefont
  {E.}~\bibnamefont {Solano}}, \ and\ \bibinfo {author} {\bibfnamefont {J.~M.}\
  \bibnamefont {Martinis}},\ }\href {\doibase 10.1038/ncomms8654} {\bibfield
  {journal} {\bibinfo  {journal} {Nature Communications}\ }\textbf {\bibinfo
  {volume} {6}},\ \bibinfo {pages} {7654} (\bibinfo {year} {2015})}\BibitemShut
  {NoStop}%
\bibitem [{\citenamefont {Schroer}\ \emph {et~al.}(2014)\citenamefont
  {Schroer}, \citenamefont {Kolodrubetz}, \citenamefont {Kindel}, \citenamefont
  {Sandberg}, \citenamefont {Gao}, \citenamefont {Vissers}, \citenamefont
  {Pappas}, \citenamefont {Polkovnikov},\ and\ \citenamefont
  {Lehnert}}]{PhysRevLett.113.050402}%
  \BibitemOpen
  \bibfield  {author} {\bibinfo {author} {\bibfnamefont {M.~D.}\ \bibnamefont
  {Schroer}}, \bibinfo {author} {\bibfnamefont {M.~H.}\ \bibnamefont
  {Kolodrubetz}}, \bibinfo {author} {\bibfnamefont {W.~F.}\ \bibnamefont
  {Kindel}}, \bibinfo {author} {\bibfnamefont {M.}~\bibnamefont {Sandberg}},
  \bibinfo {author} {\bibfnamefont {J.}~\bibnamefont {Gao}}, \bibinfo {author}
  {\bibfnamefont {M.~R.}\ \bibnamefont {Vissers}}, \bibinfo {author}
  {\bibfnamefont {D.~P.}\ \bibnamefont {Pappas}}, \bibinfo {author}
  {\bibfnamefont {A.}~\bibnamefont {Polkovnikov}}, \ and\ \bibinfo {author}
  {\bibfnamefont {K.~W.}\ \bibnamefont {Lehnert}},\ }\href {\doibase
  10.1103/PhysRevLett.113.050402} {\bibfield  {journal} {\bibinfo  {journal}
  {Phys. Rev. Lett.}\ }\textbf {\bibinfo {volume} {113}},\ \bibinfo {pages}
  {050402} (\bibinfo {year} {2014})}\BibitemShut {NoStop}%
\bibitem [{\citenamefont {Chen}\ \emph
  {et~al.}(2014{\natexlab{a}})\citenamefont {Chen}, \citenamefont {Roushan},
  \citenamefont {Sank}, \citenamefont {Neill}, \citenamefont {Lucero},
  \citenamefont {Mariantoni}, \citenamefont {Barends}, \citenamefont {Chiaro},
  \citenamefont {Kelly}, \citenamefont {Megrant}, \citenamefont {Mutus},
  \citenamefont {O'Malley}, \citenamefont {Vainsencher}, \citenamefont
  {Wenner}, \citenamefont {White}, \citenamefont {Yin}, \citenamefont
  {Cleland},\ and\ \citenamefont {Martinis}}]{Chen:2014aa}%
  \BibitemOpen
  \bibfield  {author} {\bibinfo {author} {\bibfnamefont {Y.}~\bibnamefont
  {Chen}}, \bibinfo {author} {\bibfnamefont {P.}~\bibnamefont {Roushan}},
  \bibinfo {author} {\bibfnamefont {D.}~\bibnamefont {Sank}}, \bibinfo {author}
  {\bibfnamefont {C.}~\bibnamefont {Neill}}, \bibinfo {author} {\bibfnamefont
  {E.}~\bibnamefont {Lucero}}, \bibinfo {author} {\bibfnamefont
  {M.}~\bibnamefont {Mariantoni}}, \bibinfo {author} {\bibfnamefont
  {R.}~\bibnamefont {Barends}}, \bibinfo {author} {\bibfnamefont
  {B.}~\bibnamefont {Chiaro}}, \bibinfo {author} {\bibfnamefont
  {J.}~\bibnamefont {Kelly}}, \bibinfo {author} {\bibfnamefont
  {A.}~\bibnamefont {Megrant}}, \bibinfo {author} {\bibfnamefont {J.~Y.}\
  \bibnamefont {Mutus}}, \bibinfo {author} {\bibfnamefont {P.~J.~J.}\
  \bibnamefont {O'Malley}}, \bibinfo {author} {\bibfnamefont {A.}~\bibnamefont
  {Vainsencher}}, \bibinfo {author} {\bibfnamefont {J.}~\bibnamefont {Wenner}},
  \bibinfo {author} {\bibfnamefont {T.~C.}\ \bibnamefont {White}}, \bibinfo
  {author} {\bibfnamefont {Y.}~\bibnamefont {Yin}}, \bibinfo {author}
  {\bibfnamefont {A.~N.}\ \bibnamefont {Cleland}}, \ and\ \bibinfo {author}
  {\bibfnamefont {J.~M.}\ \bibnamefont {Martinis}},\ }\href {\doibase
  10.1038/ncomms6184} {\bibfield  {journal} {\bibinfo  {journal} {Nature
  Communications}\ }\textbf {\bibinfo {volume} {5}},\ \bibinfo {pages} {5184}
  (\bibinfo {year} {2014}{\natexlab{a}})}\BibitemShut {NoStop}%
\bibitem [{\citenamefont {Roushan}\ \emph {et~al.}(2014)\citenamefont
  {Roushan}, \citenamefont {Neill}, \citenamefont {Chen}, \citenamefont
  {Kolodrubetz}, \citenamefont {Quintana}, \citenamefont {Leung}, \citenamefont
  {Fang}, \citenamefont {Barends}, \citenamefont {Campbell}, \citenamefont
  {Chen}, \citenamefont {Chiaro}, \citenamefont {Dunsworth}, \citenamefont
  {Jeffrey}, \citenamefont {Kelly}, \citenamefont {Megrant}, \citenamefont
  {Mutus}, \citenamefont {O'Malley}, \citenamefont {Sank}, \citenamefont
  {Vainsencher}, \citenamefont {Wenner}, \citenamefont {White}, \citenamefont
  {Polkovnikov}, \citenamefont {Cleland},\ and\ \citenamefont
  {Martinis}}]{Roushan:2014aa}%
  \BibitemOpen
  \bibfield  {author} {\bibinfo {author} {\bibfnamefont {P.}~\bibnamefont
  {Roushan}}, \bibinfo {author} {\bibfnamefont {C.}~\bibnamefont {Neill}},
  \bibinfo {author} {\bibfnamefont {Y.}~\bibnamefont {Chen}}, \bibinfo {author}
  {\bibfnamefont {M.}~\bibnamefont {Kolodrubetz}}, \bibinfo {author}
  {\bibfnamefont {C.}~\bibnamefont {Quintana}}, \bibinfo {author}
  {\bibfnamefont {N.}~\bibnamefont {Leung}}, \bibinfo {author} {\bibfnamefont
  {M.}~\bibnamefont {Fang}}, \bibinfo {author} {\bibfnamefont {R.}~\bibnamefont
  {Barends}}, \bibinfo {author} {\bibfnamefont {B.}~\bibnamefont {Campbell}},
  \bibinfo {author} {\bibfnamefont {Z.}~\bibnamefont {Chen}}, \bibinfo {author}
  {\bibfnamefont {B.}~\bibnamefont {Chiaro}}, \bibinfo {author} {\bibfnamefont
  {A.}~\bibnamefont {Dunsworth}}, \bibinfo {author} {\bibfnamefont
  {E.}~\bibnamefont {Jeffrey}}, \bibinfo {author} {\bibfnamefont
  {J.}~\bibnamefont {Kelly}}, \bibinfo {author} {\bibfnamefont
  {A.}~\bibnamefont {Megrant}}, \bibinfo {author} {\bibfnamefont
  {J.}~\bibnamefont {Mutus}}, \bibinfo {author} {\bibfnamefont {P.~J.~J.}\
  \bibnamefont {O'Malley}}, \bibinfo {author} {\bibfnamefont {D.}~\bibnamefont
  {Sank}}, \bibinfo {author} {\bibfnamefont {A.}~\bibnamefont {Vainsencher}},
  \bibinfo {author} {\bibfnamefont {J.}~\bibnamefont {Wenner}}, \bibinfo
  {author} {\bibfnamefont {T.}~\bibnamefont {White}}, \bibinfo {author}
  {\bibfnamefont {A.}~\bibnamefont {Polkovnikov}}, \bibinfo {author}
  {\bibfnamefont {A.~N.}\ \bibnamefont {Cleland}}, \ and\ \bibinfo {author}
  {\bibfnamefont {J.~M.}\ \bibnamefont {Martinis}},\ }\href {\doibase
  10.1038/nature13891} {\bibfield  {journal} {\bibinfo  {journal} {Nature}\
  }\textbf {\bibinfo {volume} {515}},\ \bibinfo {pages} {241} (\bibinfo {year}
  {2014})}\BibitemShut {NoStop}%
\bibitem [{\citenamefont {Besedin}\ \emph {et~al.}(2021)\citenamefont
  {Besedin}, \citenamefont {Gorlach}, \citenamefont {Abramov}, \citenamefont
  {Tsitsilin}, \citenamefont {Moskalenko}, \citenamefont {Dobronosova},
  \citenamefont {Moskalev}, \citenamefont {Matanin}, \citenamefont {Smirnov},
  \citenamefont {Rodionov}, \citenamefont {Poddubny},\ and\ \citenamefont
  {Ustinov}}]{besedin2021}%
  \BibitemOpen
  \bibfield  {author} {\bibinfo {author} {\bibfnamefont {I.~S.}\ \bibnamefont
  {Besedin}}, \bibinfo {author} {\bibfnamefont {M.~A.}\ \bibnamefont
  {Gorlach}}, \bibinfo {author} {\bibfnamefont {N.~N.}\ \bibnamefont
  {Abramov}}, \bibinfo {author} {\bibfnamefont {I.}~\bibnamefont {Tsitsilin}},
  \bibinfo {author} {\bibfnamefont {I.~N.}\ \bibnamefont {Moskalenko}},
  \bibinfo {author} {\bibfnamefont {A.~A.}\ \bibnamefont {Dobronosova}},
  \bibinfo {author} {\bibfnamefont {D.~O.}\ \bibnamefont {Moskalev}}, \bibinfo
  {author} {\bibfnamefont {A.~R.}\ \bibnamefont {Matanin}}, \bibinfo {author}
  {\bibfnamefont {N.~S.}\ \bibnamefont {Smirnov}}, \bibinfo {author}
  {\bibfnamefont {I.~A.}\ \bibnamefont {Rodionov}}, \bibinfo {author}
  {\bibfnamefont {A.~N.}\ \bibnamefont {Poddubny}}, \ and\ \bibinfo {author}
  {\bibfnamefont {A.~V.}\ \bibnamefont {Ustinov}},\ }\href {\doibase
  10.1103/PhysRevB.103.224520} {\bibfield  {journal} {\bibinfo  {journal}
  {Phys. Rev. B}\ }\textbf {\bibinfo {volume} {103}},\ \bibinfo {pages}
  {224520} (\bibinfo {year} {2021})}\BibitemShut {NoStop}%
\bibitem [{\citenamefont {Murta}\ \emph {et~al.}(2020)\citenamefont {Murta},
  \citenamefont {Catarina},\ and\ \citenamefont
  {Fern\'andez-Rossier}}]{SSH-Fernandez-Rossier}%
  \BibitemOpen
  \bibfield  {author} {\bibinfo {author} {\bibfnamefont {B.}~\bibnamefont
  {Murta}}, \bibinfo {author} {\bibfnamefont {G.}~\bibnamefont {Catarina}}, \
  and\ \bibinfo {author} {\bibfnamefont {J.}~\bibnamefont
  {Fern\'andez-Rossier}},\ }\href {\doibase 10.1103/PhysRevA.101.020302}
  {\bibfield  {journal} {\bibinfo  {journal} {Phys. Rev. A}\ }\textbf {\bibinfo
  {volume} {101}},\ \bibinfo {pages} {020302(R)} (\bibinfo {year}
  {2020})}\BibitemShut {NoStop}%
\bibitem [{\citenamefont {Hoffman}\ \emph {et~al.}(2011)\citenamefont
  {Hoffman}, \citenamefont {Srinivasan}, \citenamefont {Schmidt}, \citenamefont
  {Spietz}, \citenamefont {Aumentado}, \citenamefont {T\"ureci},\ and\
  \citenamefont {Houck}}]{PhysRevLett.107.053602}%
  \BibitemOpen
  \bibfield  {author} {\bibinfo {author} {\bibfnamefont {A.~J.}\ \bibnamefont
  {Hoffman}}, \bibinfo {author} {\bibfnamefont {S.~J.}\ \bibnamefont
  {Srinivasan}}, \bibinfo {author} {\bibfnamefont {S.}~\bibnamefont {Schmidt}},
  \bibinfo {author} {\bibfnamefont {L.}~\bibnamefont {Spietz}}, \bibinfo
  {author} {\bibfnamefont {J.}~\bibnamefont {Aumentado}}, \bibinfo {author}
  {\bibfnamefont {H.~E.}\ \bibnamefont {T\"ureci}}, \ and\ \bibinfo {author}
  {\bibfnamefont {A.~A.}\ \bibnamefont {Houck}},\ }\href {\doibase
  10.1103/PhysRevLett.107.053602} {\bibfield  {journal} {\bibinfo  {journal}
  {Phys. Rev. Lett.}\ }\textbf {\bibinfo {volume} {107}},\ \bibinfo {pages}
  {053602} (\bibinfo {year} {2011})}\BibitemShut {NoStop}%
\bibitem [{\citenamefont {Lang}\ \emph {et~al.}(2011)\citenamefont {Lang},
  \citenamefont {Bozyigit}, \citenamefont {Eichler}, \citenamefont {Steffen},
  \citenamefont {Fink}, \citenamefont {Abdumalikov}, \citenamefont {Baur},
  \citenamefont {Filipp}, \citenamefont {da~Silva}, \citenamefont {Blais},\
  and\ \citenamefont {Wallraff}}]{Photon_Blockade_corr}%
  \BibitemOpen
  \bibfield  {author} {\bibinfo {author} {\bibfnamefont {C.}~\bibnamefont
  {Lang}}, \bibinfo {author} {\bibfnamefont {D.}~\bibnamefont {Bozyigit}},
  \bibinfo {author} {\bibfnamefont {C.}~\bibnamefont {Eichler}}, \bibinfo
  {author} {\bibfnamefont {L.}~\bibnamefont {Steffen}}, \bibinfo {author}
  {\bibfnamefont {J.~M.}\ \bibnamefont {Fink}}, \bibinfo {author}
  {\bibfnamefont {A.~A.}\ \bibnamefont {Abdumalikov}}, \bibinfo {author}
  {\bibfnamefont {M.}~\bibnamefont {Baur}}, \bibinfo {author} {\bibfnamefont
  {S.}~\bibnamefont {Filipp}}, \bibinfo {author} {\bibfnamefont {M.~P.}\
  \bibnamefont {da~Silva}}, \bibinfo {author} {\bibfnamefont {A.}~\bibnamefont
  {Blais}}, \ and\ \bibinfo {author} {\bibfnamefont {A.}~\bibnamefont
  {Wallraff}},\ }\href {\doibase 10.1103/PhysRevLett.106.243601} {\bibfield
  {journal} {\bibinfo  {journal} {Phys. Rev. Lett.}\ }\textbf {\bibinfo
  {volume} {106}},\ \bibinfo {pages} {243601} (\bibinfo {year}
  {2011})}\BibitemShut {NoStop}%
\bibitem [{\citenamefont {Goetz}\ \emph {et~al.}(2017)\citenamefont {Goetz},
  \citenamefont {Pogorzalek}, \citenamefont {Deppe}, \citenamefont {Fedorov},
  \citenamefont {Eder}, \citenamefont {Fischer}, \citenamefont {Wulschner},
  \citenamefont {Xie}, \citenamefont {Marx},\ and\ \citenamefont
  {Gross}}]{Goetz:2017aa}%
  \BibitemOpen
  \bibfield  {author} {\bibinfo {author} {\bibfnamefont {J.}~\bibnamefont
  {Goetz}}, \bibinfo {author} {\bibfnamefont {S.}~\bibnamefont {Pogorzalek}},
  \bibinfo {author} {\bibfnamefont {F.}~\bibnamefont {Deppe}}, \bibinfo
  {author} {\bibfnamefont {K.~G.}\ \bibnamefont {Fedorov}}, \bibinfo {author}
  {\bibfnamefont {P.}~\bibnamefont {Eder}}, \bibinfo {author} {\bibfnamefont
  {M.}~\bibnamefont {Fischer}}, \bibinfo {author} {\bibfnamefont
  {F.}~\bibnamefont {Wulschner}}, \bibinfo {author} {\bibfnamefont
  {E.}~\bibnamefont {Xie}}, \bibinfo {author} {\bibfnamefont {A.}~\bibnamefont
  {Marx}}, \ and\ \bibinfo {author} {\bibfnamefont {R.}~\bibnamefont {Gross}},\
  }\href {\doibase 10.1103/PhysRevLett.118.103602} {\bibfield  {journal}
  {\bibinfo  {journal} {Phys. Rev. Lett.}\ }\textbf {\bibinfo {volume} {118}},\
  \bibinfo {pages} {103602} (\bibinfo {year} {2017})}\BibitemShut {NoStop}%
\bibitem [{\citenamefont {Dmitriev}\ \emph {et~al.}(2019)\citenamefont
  {Dmitriev}, \citenamefont {Shaikhaidarov}, \citenamefont {H\"onigl-Decrinis},
  \citenamefont {de~Graaf}, \citenamefont {Antonov},\ and\ \citenamefont
  {Astafiev}}]{Dmitriev:2019aa}%
  \BibitemOpen
  \bibfield  {author} {\bibinfo {author} {\bibfnamefont {A.~Y.}\ \bibnamefont
  {Dmitriev}}, \bibinfo {author} {\bibfnamefont {R.}~\bibnamefont
  {Shaikhaidarov}}, \bibinfo {author} {\bibfnamefont {T.}~\bibnamefont
  {H\"onigl-Decrinis}}, \bibinfo {author} {\bibfnamefont {S.~E.}\ \bibnamefont
  {de~Graaf}}, \bibinfo {author} {\bibfnamefont {V.~N.}\ \bibnamefont
  {Antonov}}, \ and\ \bibinfo {author} {\bibfnamefont {O.~V.}\ \bibnamefont
  {Astafiev}},\ }\href {\doibase 10.1103/PhysRevA.100.013808} {\bibfield
  {journal} {\bibinfo  {journal} {Phys. Rev. A}\ }\textbf {\bibinfo {volume}
  {100}},\ \bibinfo {pages} {013808} (\bibinfo {year} {2019})}\BibitemShut
  {NoStop}%
\bibitem [{\citenamefont {H\"onigl-Decrinis}\ \emph {et~al.}(2020)\citenamefont
  {H\"onigl-Decrinis}, \citenamefont {Shaikhaidarov}, \citenamefont {de~Graaf},
  \citenamefont {Antonov},\ and\ \citenamefont
  {Astafiev}}]{Honigl-Decrinis:2020aa}%
  \BibitemOpen
  \bibfield  {author} {\bibinfo {author} {\bibfnamefont {T.}~\bibnamefont
  {H\"onigl-Decrinis}}, \bibinfo {author} {\bibfnamefont {R.}~\bibnamefont
  {Shaikhaidarov}}, \bibinfo {author} {\bibfnamefont {S.~E.}~\bibnamefont
  {de~Graaf}}, \bibinfo {author} {\bibfnamefont {V.~N.}~\bibnamefont {Antonov}}, \
  and\ \bibinfo {author} {\bibfnamefont {O.~V.}~\bibnamefont {Astafiev}},\ }\href
  {\doibase 10.1103/PhysRevApplied.13.024066} {\bibfield  {journal} {\bibinfo
  {journal} {Phys. Rev. Applied}\ }\textbf {\bibinfo {volume} {13}},\ \bibinfo
  {pages} {024066} (\bibinfo {year} {2020})}\BibitemShut {NoStop}%
\bibitem [{\citenamefont {Zhou}\ \emph {et~al.}(2020)\citenamefont {Zhou},
  \citenamefont {Peng}, \citenamefont {Horiuchi}, \citenamefont {Astafiev},\
  and\ \citenamefont {Tsai}}]{Zhou:2020aa}%
  \BibitemOpen
  \bibfield  {author} {\bibinfo {author} {\bibfnamefont {Y.}~\bibnamefont
  {Zhou}}, \bibinfo {author} {\bibfnamefont {Z.}~\bibnamefont {Peng}}, \bibinfo
  {author} {\bibfnamefont {Y.}~\bibnamefont {Horiuchi}}, \bibinfo {author}
  {\bibfnamefont {O.~V.}~\bibnamefont {Astafiev}}, \ and\ \bibinfo {author}
  {\bibfnamefont {J.~S.}~\bibnamefont {Tsai}},\ }\href {\doibase
  10.1103/PhysRevApplied.13.034007} {\bibfield  {journal} {\bibinfo  {journal}
  {Phys. Rev. Applied}\ }\textbf {\bibinfo {volume} {13}},\ \bibinfo {pages}
  {034007} (\bibinfo {year} {2020})}\BibitemShut {NoStop}%
\bibitem [{\citenamefont {Braum\"uller}\ \emph {et~al.}(2015)\citenamefont
  {Braum\"uller}, \citenamefont {Cramer}, \citenamefont {Schl\"or},
  \citenamefont {Rotzinger}, \citenamefont {Radtke}, \citenamefont
  {Lukashenko}, \citenamefont {Yang}, \citenamefont {Skacel}, \citenamefont
  {Probst}, \citenamefont {Marthaler}, \citenamefont {Guo}, \citenamefont
  {Ustinov},\ and\ \citenamefont {Weides}}]{Braumueller2015}%
  \BibitemOpen
  \bibfield  {author} {\bibinfo {author} {\bibfnamefont {J.}~\bibnamefont
  {Braum\"uller}}, \bibinfo {author} {\bibfnamefont {J.}~\bibnamefont
  {Cramer}}, \bibinfo {author} {\bibfnamefont {S.}~\bibnamefont {Schl\"or}},
  \bibinfo {author} {\bibfnamefont {H.}~\bibnamefont {Rotzinger}}, \bibinfo
  {author} {\bibfnamefont {L.}~\bibnamefont {Radtke}}, \bibinfo {author}
  {\bibfnamefont {A.}~\bibnamefont {Lukashenko}}, \bibinfo {author}
  {\bibfnamefont {P.}~\bibnamefont {Yang}}, \bibinfo {author} {\bibfnamefont
  {S.~T.}\ \bibnamefont {Skacel}}, \bibinfo {author} {\bibfnamefont
  {S.}~\bibnamefont {Probst}}, \bibinfo {author} {\bibfnamefont
  {M.}~\bibnamefont {Marthaler}}, \bibinfo {author} {\bibfnamefont
  {L.}~\bibnamefont {Guo}}, \bibinfo {author} {\bibfnamefont {A.~V.}\
  \bibnamefont {Ustinov}}, \ and\ \bibinfo {author} {\bibfnamefont
  {M.}~\bibnamefont {Weides}},\ }\href {\doibase 10.1103/PhysRevB.91.054523}
  {\bibfield  {journal} {\bibinfo  {journal} {Phys. Rev. B}\ }\textbf {\bibinfo
  {volume} {91}},\ \bibinfo {pages} {054523} (\bibinfo {year}
  {2015})}\BibitemShut {NoStop}%
\bibitem [{\citenamefont {Macha}\ \emph {et~al.}(2014)\citenamefont {Macha},
  \citenamefont {Oelsner}, \citenamefont {Reiner}, \citenamefont {Marthaler},
  \citenamefont {Andr{\'e}}, \citenamefont {Sch{\"o}n}, \citenamefont
  {H{\"u}bner}, \citenamefont {Meyer}, \citenamefont {Il'ichev},\ and\
  \citenamefont {Ustinov}}]{macha2014implementation}%
  \BibitemOpen
  \bibfield  {author} {\bibinfo {author} {\bibfnamefont {P.}~\bibnamefont
  {Macha}}, \bibinfo {author} {\bibfnamefont {G.}~\bibnamefont {Oelsner}},
  \bibinfo {author} {\bibfnamefont {J.-M.}\ \bibnamefont {Reiner}}, \bibinfo
  {author} {\bibfnamefont {M.}~\bibnamefont {Marthaler}}, \bibinfo {author}
  {\bibfnamefont {S.}~\bibnamefont {Andr{\'e}}}, \bibinfo {author}
  {\bibfnamefont {G.}~\bibnamefont {Sch{\"o}n}}, \bibinfo {author}
  {\bibfnamefont {U.}~\bibnamefont {H{\"u}bner}}, \bibinfo {author}
  {\bibfnamefont {H.-G.}\ \bibnamefont {Meyer}}, \bibinfo {author}
  {\bibfnamefont {E.}~\bibnamefont {Il'ichev}}, \ and\ \bibinfo {author}
  {\bibfnamefont {A.~V.}\ \bibnamefont {Ustinov}},\ }\href@noop {} {\bibfield
  {journal} {\bibinfo  {journal} {Nature communications}\ }\textbf {\bibinfo
  {volume} {5}},\ \bibinfo {pages} {5146} (\bibinfo {year} {2014})}\BibitemShut
  {NoStop}%
\bibitem [{\citenamefont {Frisk~Kockum}\ \emph {et~al.}(2019)\citenamefont
  {Frisk~Kockum}, \citenamefont {Miranowicz}, \citenamefont {De~Liberato},
  \citenamefont {Savasta},\ and\ \citenamefont {Nori}}]{Frisk-Kockum:2019aa}%
  \BibitemOpen
  \bibfield  {author} {\bibinfo {author} {\bibfnamefont {A.}~\bibnamefont
  {Frisk~Kockum}}, \bibinfo {author} {\bibfnamefont {A.}~\bibnamefont
  {Miranowicz}}, \bibinfo {author} {\bibfnamefont {S.}~\bibnamefont
  {De~Liberato}}, \bibinfo {author} {\bibfnamefont {S.}~\bibnamefont
  {Savasta}}, \ and\ \bibinfo {author} {\bibfnamefont {F.}~\bibnamefont
  {Nori}},\ }\href {\doibase 10.1038/s42254-018-0006-2} {\bibfield  {journal}
  {\bibinfo  {journal} {Nature Reviews Physics}\ }\textbf {\bibinfo {volume}
  {1}},\ \bibinfo {pages} {19} (\bibinfo {year} {2019})}\BibitemShut {NoStop}%
\bibitem [{\citenamefont {Kirton}\ \emph {et~al.}(2019)\citenamefont {Kirton},
  \citenamefont {Roses}, \citenamefont {Keeling},\ and\ \citenamefont {{Dalla
  Torre}}}]{kirton2018introduction}%
  \BibitemOpen
  \bibfield  {author} {\bibinfo {author} {\bibfnamefont {P.}~\bibnamefont
  {Kirton}}, \bibinfo {author} {\bibfnamefont {M.~M.}\ \bibnamefont {Roses}},
  \bibinfo {author} {\bibfnamefont {J.}~\bibnamefont {Keeling}}, \ and\
  \bibinfo {author} {\bibfnamefont {E.~G.}\ \bibnamefont {{Dalla Torre}}},\
  }\href@noop {} {\bibfield  {journal} {\bibinfo  {journal} {Adv. Quantum
  Technol.}\ }\textbf {\bibinfo {volume} {2}},\ \bibinfo {pages} {1800043}
  (\bibinfo {year} {2019})}\BibitemShut {NoStop}%
\bibitem [{\citenamefont {Shapiro}\ \emph {et~al.}(2020)\citenamefont
  {Shapiro}, \citenamefont {Pogosov},\ and\ \citenamefont
  {Lozovik}}]{Shapiro2020}%
  \BibitemOpen
  \bibfield  {author} {\bibinfo {author} {\bibfnamefont {D.~S.}\ \bibnamefont
  {Shapiro}}, \bibinfo {author} {\bibfnamefont {W.~V.}\ \bibnamefont
  {Pogosov}}, \ and\ \bibinfo {author} {\bibfnamefont {Y.~E.}\ \bibnamefont
  {Lozovik}},\ }\href {\doibase 10.1103/PhysRevA.102.023703} {\bibfield
  {journal} {\bibinfo  {journal} {Phys. Rev. A}\ }\textbf {\bibinfo {volume}
  {102}},\ \bibinfo {pages} {023703} (\bibinfo {year} {2020})}\BibitemShut
  {NoStop}%
\bibitem [{\citenamefont {Biella}\ \emph {et~al.}(2015)\citenamefont {Biella},
  \citenamefont {Mazza}, \citenamefont {Carusotto}, \citenamefont {Rossini},\
  and\ \citenamefont {Fazio}}]{Biella:2015aa}%
  \BibitemOpen
  \bibfield  {author} {\bibinfo {author} {\bibfnamefont {A.}~\bibnamefont
  {Biella}}, \bibinfo {author} {\bibfnamefont {L.}~\bibnamefont {Mazza}},
  \bibinfo {author} {\bibfnamefont {I.}~\bibnamefont {Carusotto}}, \bibinfo
  {author} {\bibfnamefont {D.}~\bibnamefont {Rossini}}, \ and\ \bibinfo
  {author} {\bibfnamefont {R.}~\bibnamefont {Fazio}},\ }\href {\doibase
  10.1103/PhysRevA.91.053815} {\bibfield  {journal} {\bibinfo  {journal} {Phys.
  Rev. A}\ }\textbf {\bibinfo {volume} {91}},\ \bibinfo {pages} {053815}
  (\bibinfo {year} {2015})}\BibitemShut {NoStop}%
\bibitem [{\citenamefont {Vicentini}\ \emph {et~al.}(2018)\citenamefont
  {Vicentini}, \citenamefont {Minganti}, \citenamefont {Rota}, \citenamefont
  {Orso},\ and\ \citenamefont {Ciuti}}]{Vicentini:2018aa}%
  \BibitemOpen
  \bibfield  {author} {\bibinfo {author} {\bibfnamefont {F.}~\bibnamefont
  {Vicentini}}, \bibinfo {author} {\bibfnamefont {F.}~\bibnamefont {Minganti}},
  \bibinfo {author} {\bibfnamefont {R.}~\bibnamefont {Rota}}, \bibinfo {author}
  {\bibfnamefont {G.}~\bibnamefont {Orso}}, \ and\ \bibinfo {author}
  {\bibfnamefont {C.}~\bibnamefont {Ciuti}},\ }\href {\doibase
  10.1103/PhysRevA.97.013853} {\bibfield  {journal} {\bibinfo  {journal} {Phys.
  Rev. A}\ }\textbf {\bibinfo {volume} {97}},\ \bibinfo {pages} {013853}
  (\bibinfo {year} {2018})}\BibitemShut {NoStop}%
\bibitem [{\citenamefont {Fedorov}\ \emph {et~al.}(2021)\citenamefont
  {Fedorov}, \citenamefont {Remizov}, \citenamefont {Shapiro}, \citenamefont
  {Pogosov}, \citenamefont {Egorova}, \citenamefont {Tsitsilin}, \citenamefont
  {Andronik}, \citenamefont {Dobronosova}, \citenamefont {Rodionov},
  \citenamefont {Astafiev},\ and\ \citenamefont {Ustinov}}]{Gleb-PRL-2021}%
  \BibitemOpen
  \bibfield  {author} {\bibinfo {author} {\bibfnamefont {G.~P.}\ \bibnamefont
  {Fedorov}}, \bibinfo {author} {\bibfnamefont {S.~V.}\ \bibnamefont
  {Remizov}}, \bibinfo {author} {\bibfnamefont {D.~S.}\ \bibnamefont
  {Shapiro}}, \bibinfo {author} {\bibfnamefont {W.~V.}\ \bibnamefont
  {Pogosov}}, \bibinfo {author} {\bibfnamefont {E.}~\bibnamefont {Egorova}},
  \bibinfo {author} {\bibfnamefont {I.}~\bibnamefont {Tsitsilin}}, \bibinfo
  {author} {\bibfnamefont {M.}~\bibnamefont {Andronik}}, \bibinfo {author}
  {\bibfnamefont {A.~A.}\ \bibnamefont {Dobronosova}}, \bibinfo {author}
  {\bibfnamefont {I.~A.}\ \bibnamefont {Rodionov}}, \bibinfo {author}
  {\bibfnamefont {O.~V.}\ \bibnamefont {Astafiev}}, \ and\ \bibinfo {author}
  {\bibfnamefont {A.~V.}\ \bibnamefont {Ustinov}},\ }\href {\doibase
  10.1103/PhysRevLett.126.180503} {\bibfield  {journal} {\bibinfo  {journal}
  {Phys. Rev. Lett.}\ }\textbf {\bibinfo {volume} {126}},\ \bibinfo {pages}
  {180503} (\bibinfo {year} {2021})}\BibitemShut {NoStop}%
\bibitem [{\citenamefont {Shapiro}\ \emph {et~al.}(2015)\citenamefont
  {Shapiro}, \citenamefont {Macha}, \citenamefont {Rubtsov},\ and\
  \citenamefont {Ustinov}}]{Shapiro2015}%
  \BibitemOpen
  \bibfield  {author} {\bibinfo {author} {\bibfnamefont {D.~S.}\ \bibnamefont
  {Shapiro}}, \bibinfo {author} {\bibfnamefont {P.}~\bibnamefont {Macha}},
  \bibinfo {author} {\bibfnamefont {A.~N.}\ \bibnamefont {Rubtsov}}, \ and\
  \bibinfo {author} {\bibfnamefont {A.~V.}\ \bibnamefont {Ustinov}},\ }\href
  {\doibase 10.3390/photonics2020449} {\bibfield  {journal} {\bibinfo
  {journal} {Photonics}\ }\textbf {\bibinfo {volume} {2}},\ \bibinfo {pages}
  {449} (\bibinfo {year} {2015})}\BibitemShut {NoStop}%
\bibitem [{\citenamefont {Shulga}\ \emph {et~al.}(2017)\citenamefont {Shulga},
  \citenamefont {Yang}, \citenamefont {Fedorov}, \citenamefont {Fistul},
  \citenamefont {Weides},\ and\ \citenamefont {Ustinov}}]{Shulga2017}%
  \BibitemOpen
  \bibfield  {author} {\bibinfo {author} {\bibfnamefont {K.~V.}\ \bibnamefont
  {Shulga}}, \bibinfo {author} {\bibfnamefont {P.}~\bibnamefont {Yang}},
  \bibinfo {author} {\bibfnamefont {G.~P.}\ \bibnamefont {Fedorov}}, \bibinfo
  {author} {\bibfnamefont {M.~V.}\ \bibnamefont {Fistul}}, \bibinfo {author}
  {\bibfnamefont {M.}~\bibnamefont {Weides}}, \ and\ \bibinfo {author}
  {\bibfnamefont {A.~V.}\ \bibnamefont {Ustinov}},\ }\href {\doibase
  10.1134/S0021364017010143} {\bibfield  {journal} {\bibinfo  {journal} {JETP
  Letters}\ }\textbf {\bibinfo {volume} {105}},\ \bibinfo {pages} {47}
  (\bibinfo {year} {2017})}\BibitemShut {NoStop}%
\bibitem [{\citenamefont {Tavis}\ and\ \citenamefont
  {Cummings}(1968)}]{Tav-Cumm-1968}%
  \BibitemOpen
  \bibfield  {author} {\bibinfo {author} {\bibfnamefont {M.}~\bibnamefont
  {Tavis}}\ and\ \bibinfo {author} {\bibfnamefont {F.~W.}\ \bibnamefont
  {Cummings}},\ }\href {\doibase 10.1103/PhysRev.170.379} {\bibfield  {journal}
  {\bibinfo  {journal} {Phys. Rev.}\ }\textbf {\bibinfo {volume} {170}},\
  \bibinfo {pages} {379} (\bibinfo {year} {1968})}\BibitemShut {NoStop}%
\bibitem [{\citenamefont {Koch}\ \emph {et~al.}(2007)\citenamefont {Koch},
  \citenamefont {Yu}, \citenamefont {Gambetta}, \citenamefont {Houck},
  \citenamefont {Schuster}, \citenamefont {Majer}, \citenamefont {Blais},
  \citenamefont {Devoret}, \citenamefont {Girvin},\ and\ \citenamefont
  {Schoelkopf}}]{PhysRevA.76.042319}%
  \BibitemOpen
  \bibfield  {author} {\bibinfo {author} {\bibfnamefont {J.}~\bibnamefont
  {Koch}}, \bibinfo {author} {\bibfnamefont {T.~M.}\ \bibnamefont {Yu}},
  \bibinfo {author} {\bibfnamefont {J.}~\bibnamefont {Gambetta}}, \bibinfo
  {author} {\bibfnamefont {A.~A.}\ \bibnamefont {Houck}}, \bibinfo {author}
  {\bibfnamefont {D.~I.}\ \bibnamefont {Schuster}}, \bibinfo {author}
  {\bibfnamefont {J.}~\bibnamefont {Majer}}, \bibinfo {author} {\bibfnamefont
  {A.}~\bibnamefont {Blais}}, \bibinfo {author} {\bibfnamefont {M.~H.}\
  \bibnamefont {Devoret}}, \bibinfo {author} {\bibfnamefont {S.~M.}\
  \bibnamefont {Girvin}}, \ and\ \bibinfo {author} {\bibfnamefont {R.~J.}\
  \bibnamefont {Schoelkopf}},\ }\href {\doibase 10.1103/PhysRevA.76.042319}
  {\bibfield  {journal} {\bibinfo  {journal} {Phys. Rev. A}\ }\textbf {\bibinfo
  {volume} {76}},\ \bibinfo {pages} {042319} (\bibinfo {year}
  {2007})}\BibitemShut {NoStop}%
\bibitem [{\citenamefont {Smith}\ \emph {et~al.}(2019)\citenamefont {Smith},
  \citenamefont {Kim}, \citenamefont {Pollmann},\ and\ \citenamefont
  {Knolle}}]{Smith:2019aa}%
  \BibitemOpen
  \bibfield  {author} {\bibinfo {author} {\bibfnamefont {A.}~\bibnamefont
  {Smith}}, \bibinfo {author} {\bibfnamefont {M.~S.}\ \bibnamefont {Kim}},
  \bibinfo {author} {\bibfnamefont {F.}~\bibnamefont {Pollmann}}, \ and\
  \bibinfo {author} {\bibfnamefont {J.}~\bibnamefont {Knolle}},\ }\href
  {\doibase 10.1038/s41534-019-0217-0} {\bibfield  {journal} {\bibinfo
  {journal} {npj Quantum Information}\ }\textbf {\bibinfo {volume} {5}},\
  \bibinfo {pages} {106} (\bibinfo {year} {2019})}\BibitemShut {NoStop}%
\bibitem [{\citenamefont {Braum\"uller}\ \emph {et~al.}(2020)\citenamefont
  {Braum\"uller}, \citenamefont {Ding}, \citenamefont {Veps\"al\"ainen},
  \citenamefont {Sung}, \citenamefont {Kjaergaard}, \citenamefont {Menke},
  \citenamefont {Winik}, \citenamefont {Kim}, \citenamefont {Niedzielski},
  \citenamefont {Melville}, \citenamefont {Yoder}, \citenamefont
  {Hirjibehedin}, \citenamefont {Orlando}, \citenamefont {Gustavsson},\ and\
  \citenamefont {Oliver}}]{PhysRevApplied.13.054079}%
  \BibitemOpen
  \bibfield  {author} {\bibinfo {author} {\bibfnamefont {J.}~\bibnamefont
  {Braum\"uller}}, \bibinfo {author} {\bibfnamefont {L.}~\bibnamefont {Ding}},
  \bibinfo {author} {\bibfnamefont {A.~P.}\ \bibnamefont {Veps\"al\"ainen}},
  \bibinfo {author} {\bibfnamefont {Y.}~\bibnamefont {Sung}}, \bibinfo {author}
  {\bibfnamefont {M.}~\bibnamefont {Kjaergaard}}, \bibinfo {author}
  {\bibfnamefont {T.}~\bibnamefont {Menke}}, \bibinfo {author} {\bibfnamefont
  {R.}~\bibnamefont {Winik}}, \bibinfo {author} {\bibfnamefont
  {D.}~\bibnamefont {Kim}}, \bibinfo {author} {\bibfnamefont {B.~M.}\
  \bibnamefont {Niedzielski}}, \bibinfo {author} {\bibfnamefont
  {A.}~\bibnamefont {Melville}}, \bibinfo {author} {\bibfnamefont {J.~L.}\
  \bibnamefont {Yoder}}, \bibinfo {author} {\bibfnamefont {C.~F.}\ \bibnamefont
  {Hirjibehedin}}, \bibinfo {author} {\bibfnamefont {T.~P.}\ \bibnamefont
  {Orlando}}, \bibinfo {author} {\bibfnamefont {S.}~\bibnamefont {Gustavsson}},
  \ and\ \bibinfo {author} {\bibfnamefont {W.~D.}\ \bibnamefont {Oliver}},\
  }\href {\doibase 10.1103/PhysRevApplied.13.054079} {\bibfield  {journal}
  {\bibinfo  {journal} {Phys. Rev. Applied}\ }\textbf {\bibinfo {volume}
  {13}},\ \bibinfo {pages} {054079} (\bibinfo {year} {2020})}\BibitemShut
  {NoStop}%
\bibitem [{\citenamefont {Arute}\ \emph {et~al.}(2020)\citenamefont {Arute},
  \citenamefont {Arya}, \citenamefont {Babbush}, \citenamefont {Bacon},
  \citenamefont {Bardin}, \citenamefont {Barends}, \citenamefont {Bengtsson},
  \citenamefont {Boixo}, \citenamefont {Broughton}, \citenamefont {Buckley},
  \citenamefont {Buell}, \citenamefont {Burkett}, \citenamefont {Bushnell},
  \citenamefont {Chen}, \citenamefont {Chen}, \citenamefont {Chen},
  \citenamefont {Chiaro}, \citenamefont {Collins}, \citenamefont {Cotton},
  \citenamefont {Courtney}, \citenamefont {Demura}, \citenamefont {Derk},
  \citenamefont {Dunsworth}, \citenamefont {Eppens}, \citenamefont {Eckl},
  \citenamefont {Erickson}, \citenamefont {Farhi}, \citenamefont {Fowler},
  \citenamefont {Foxen}, \citenamefont {Gidney}, \citenamefont {Giustina},
  \citenamefont {Graff}, \citenamefont {Gross}, \citenamefont {Habegger},
  \citenamefont {Harrigan}, \citenamefont {Ho}, \citenamefont {Hong},
  \citenamefont {Huang}, \citenamefont {Huggins}, \citenamefont {Ioffe},
  \citenamefont {Isakov}, \citenamefont {Jeffrey}, \citenamefont {Jiang},
  \citenamefont {Jones}, \citenamefont {Kafri}, \citenamefont {Kechedzhi},
  \citenamefont {Kelly}, \citenamefont {Kim}, \citenamefont {Klimov},
  \citenamefont {Korotkov}, \citenamefont {Kostritsa}, \citenamefont
  {Landhuis}, \citenamefont {Laptev}, \citenamefont {Lindmark}, \citenamefont
  {Lucero}, \citenamefont {Marthaler}, \citenamefont {Martin}, \citenamefont
  {Martinis}, \citenamefont {Marusczyk}, \citenamefont {McArdle}, \citenamefont
  {McClean}, \citenamefont {McCourt}, \citenamefont {McEwen}, \citenamefont
  {Megrant}, \citenamefont {Mejuto-Zaera}, \citenamefont {Mi}, \citenamefont
  {Mohseni}, \citenamefont {Mruczkiewicz}, \citenamefont {Mutus}, \citenamefont
  {Naaman}, \citenamefont {Neeley}, \citenamefont {Neill}, \citenamefont
  {Neven}, \citenamefont {Newman}, \citenamefont {Niu}, \citenamefont
  {O'Brien}, \citenamefont {Ostby}, \citenamefont {Pat{\'o}}, \citenamefont
  {Petukhov}, \citenamefont {Putterman}, \citenamefont {Quintana},
  \citenamefont {Reiner}, \citenamefont {Roushan}, \citenamefont {Rubin},
  \citenamefont {Sank}, \citenamefont {Satzinger}, \citenamefont {Smelyanskiy},
  \citenamefont {Strain}, \citenamefont {Sung}, \citenamefont {Schmitteckert},
  \citenamefont {Szalay}, \citenamefont {Tubman}, \citenamefont {Vainsencher},
  \citenamefont {White}, \citenamefont {Vogt}, \citenamefont {Yao},
  \citenamefont {Yeh}, \citenamefont {Zalcman},\ and\ \citenamefont
  {Zanker}}]{arute2020observation}%
  \BibitemOpen
  \bibfield  {author} {\bibinfo {author} {\bibfnamefont {F.}~\bibnamefont
  {Arute}}, \bibinfo {author} {\bibfnamefont {K.}~\bibnamefont {Arya}},
  \bibinfo {author} {\bibfnamefont {R.}~\bibnamefont {Babbush}}, \bibinfo
  {author} {\bibfnamefont {D.}~\bibnamefont {Bacon}}, \bibinfo {author}
  {\bibfnamefont {J.~C.}\ \bibnamefont {Bardin}}, \bibinfo {author}
  {\bibfnamefont {R.}~\bibnamefont {Barends}}, \bibinfo {author} {\bibfnamefont
  {A.}~\bibnamefont {Bengtsson}}, \bibinfo {author} {\bibfnamefont
  {S.}~\bibnamefont {Boixo}}, \bibinfo {author} {\bibfnamefont
  {M.}~\bibnamefont {Broughton}}, \bibinfo {author} {\bibfnamefont {B.~B.}\
  \bibnamefont {Buckley}}, \bibinfo {author} {\bibfnamefont {D.~A.}\
  \bibnamefont {Buell}}, \bibinfo {author} {\bibfnamefont {B.}~\bibnamefont
  {Burkett}}, \bibinfo {author} {\bibfnamefont {N.}~\bibnamefont {Bushnell}},
  \bibinfo {author} {\bibfnamefont {Y.}~\bibnamefont {Chen}}, \bibinfo {author}
  {\bibfnamefont {Z.}~\bibnamefont {Chen}}, \bibinfo {author} {\bibfnamefont
  {Y.-A.}\ \bibnamefont {Chen}}, \bibinfo {author} {\bibfnamefont
  {B.}~\bibnamefont {Chiaro}}, \bibinfo {author} {\bibfnamefont
  {R.}~\bibnamefont {Collins}}, \bibinfo {author} {\bibfnamefont {S.~J.}\
  \bibnamefont {Cotton}}, \bibinfo {author} {\bibfnamefont {W.}~\bibnamefont
  {Courtney}}, \bibinfo {author} {\bibfnamefont {S.}~\bibnamefont {Demura}},
  \bibinfo {author} {\bibfnamefont {A.}~\bibnamefont {Derk}}, \bibinfo {author}
  {\bibfnamefont {A.}~\bibnamefont {Dunsworth}}, \bibinfo {author}
  {\bibfnamefont {D.}~\bibnamefont {Eppens}}, \bibinfo {author} {\bibfnamefont
  {T.}~\bibnamefont {Eckl}}, \bibinfo {author} {\bibfnamefont {C.}~\bibnamefont
  {Erickson}}, \bibinfo {author} {\bibfnamefont {E.}~\bibnamefont {Farhi}},
  \bibinfo {author} {\bibfnamefont {A.}~\bibnamefont {Fowler}}, \bibinfo
  {author} {\bibfnamefont {B.}~\bibnamefont {Foxen}}, \bibinfo {author}
  {\bibfnamefont {C.}~\bibnamefont {Gidney}}, \bibinfo {author} {\bibfnamefont
  {M.}~\bibnamefont {Giustina}}, \bibinfo {author} {\bibfnamefont
  {R.}~\bibnamefont {Graff}}, \bibinfo {author} {\bibfnamefont {J.~A.}\
  \bibnamefont {Gross}}, \bibinfo {author} {\bibfnamefont {S.}~\bibnamefont
  {Habegger}}, \bibinfo {author} {\bibfnamefont {M.~P.}\ \bibnamefont
  {Harrigan}}, \bibinfo {author} {\bibfnamefont {A.}~\bibnamefont {Ho}},
  \bibinfo {author} {\bibfnamefont {S.}~\bibnamefont {Hong}}, \bibinfo {author}
  {\bibfnamefont {T.}~\bibnamefont {Huang}}, \bibinfo {author} {\bibfnamefont
  {W.}~\bibnamefont {Huggins}}, \bibinfo {author} {\bibfnamefont {L.~B.}\
  \bibnamefont {Ioffe}}, \bibinfo {author} {\bibfnamefont {S.~V.}\ \bibnamefont
  {Isakov}}, \bibinfo {author} {\bibfnamefont {E.}~\bibnamefont {Jeffrey}},
  \bibinfo {author} {\bibfnamefont {Z.}~\bibnamefont {Jiang}}, \bibinfo
  {author} {\bibfnamefont {C.}~\bibnamefont {Jones}}, \bibinfo {author}
  {\bibfnamefont {D.}~\bibnamefont {Kafri}}, \bibinfo {author} {\bibfnamefont
  {K.}~\bibnamefont {Kechedzhi}}, \bibinfo {author} {\bibfnamefont
  {J.}~\bibnamefont {Kelly}}, \bibinfo {author} {\bibfnamefont
  {S.}~\bibnamefont {Kim}}, \bibinfo {author} {\bibfnamefont {P.~V.}\
  \bibnamefont {Klimov}}, \bibinfo {author} {\bibfnamefont {A.~N.}\
  \bibnamefont {Korotkov}}, \bibinfo {author} {\bibfnamefont {F.}~\bibnamefont
  {Kostritsa}}, \bibinfo {author} {\bibfnamefont {D.}~\bibnamefont {Landhuis}},
  \bibinfo {author} {\bibfnamefont {P.}~\bibnamefont {Laptev}}, \bibinfo
  {author} {\bibfnamefont {M.}~\bibnamefont {Lindmark}}, \bibinfo {author}
  {\bibfnamefont {E.}~\bibnamefont {Lucero}}, \bibinfo {author} {\bibfnamefont
  {M.}~\bibnamefont {Marthaler}}, \bibinfo {author} {\bibfnamefont
  {O.}~\bibnamefont {Martin}}, \bibinfo {author} {\bibfnamefont {J.~M.}\
  \bibnamefont {Martinis}}, \bibinfo {author} {\bibfnamefont {A.}~\bibnamefont
  {Marusczyk}}, \bibinfo {author} {\bibfnamefont {S.}~\bibnamefont {McArdle}},
  \bibinfo {author} {\bibfnamefont {J.~R.}\ \bibnamefont {McClean}}, \bibinfo
  {author} {\bibfnamefont {T.}~\bibnamefont {McCourt}}, \bibinfo {author}
  {\bibfnamefont {M.}~\bibnamefont {McEwen}}, \bibinfo {author} {\bibfnamefont
  {A.}~\bibnamefont {Megrant}}, \bibinfo {author} {\bibfnamefont
  {C.}~\bibnamefont {Mejuto-Zaera}}, \bibinfo {author} {\bibfnamefont
  {X.}~\bibnamefont {Mi}}, \bibinfo {author} {\bibfnamefont {M.}~\bibnamefont
  {Mohseni}}, \bibinfo {author} {\bibfnamefont {W.}~\bibnamefont
  {Mruczkiewicz}}, \bibinfo {author} {\bibfnamefont {J.}~\bibnamefont {Mutus}},
  \bibinfo {author} {\bibfnamefont {O.}~\bibnamefont {Naaman}}, \bibinfo
  {author} {\bibfnamefont {M.}~\bibnamefont {Neeley}}, \bibinfo {author}
  {\bibfnamefont {C.}~\bibnamefont {Neill}}, \bibinfo {author} {\bibfnamefont
  {H.}~\bibnamefont {Neven}}, \bibinfo {author} {\bibfnamefont
  {M.}~\bibnamefont {Newman}}, \bibinfo {author} {\bibfnamefont {M.~Y.}\
  \bibnamefont {Niu}}, \bibinfo {author} {\bibfnamefont {T.~E.}\ \bibnamefont
  {O'Brien}}, \bibinfo {author} {\bibfnamefont {E.}~\bibnamefont {Ostby}},
  \bibinfo {author} {\bibfnamefont {B.}~\bibnamefont {Pat{\'o}}}, \bibinfo
  {author} {\bibfnamefont {A.}~\bibnamefont {Petukhov}}, \bibinfo {author}
  {\bibfnamefont {H.}~\bibnamefont {Putterman}}, \bibinfo {author}
  {\bibfnamefont {C.}~\bibnamefont {Quintana}}, \bibinfo {author}
  {\bibfnamefont {J.-M.}\ \bibnamefont {Reiner}}, \bibinfo {author}
  {\bibfnamefont {P.}~\bibnamefont {Roushan}}, \bibinfo {author} {\bibfnamefont
  {N.~C.}\ \bibnamefont {Rubin}}, \bibinfo {author} {\bibfnamefont
  {D.}~\bibnamefont {Sank}}, \bibinfo {author} {\bibfnamefont {K.~J.}\
  \bibnamefont {Satzinger}}, \bibinfo {author} {\bibfnamefont {V.}~\bibnamefont
  {Smelyanskiy}}, \bibinfo {author} {\bibfnamefont {D.}~\bibnamefont {Strain}},
  \bibinfo {author} {\bibfnamefont {K.~J.}\ \bibnamefont {Sung}}, \bibinfo
  {author} {\bibfnamefont {P.}~\bibnamefont {Schmitteckert}}, \bibinfo {author}
  {\bibfnamefont {M.}~\bibnamefont {Szalay}}, \bibinfo {author} {\bibfnamefont
  {N.~M.}\ \bibnamefont {Tubman}}, \bibinfo {author} {\bibfnamefont
  {A.}~\bibnamefont {Vainsencher}}, \bibinfo {author} {\bibfnamefont
  {T.}~\bibnamefont {White}}, \bibinfo {author} {\bibfnamefont
  {N.}~\bibnamefont {Vogt}}, \bibinfo {author} {\bibfnamefont {Z.~J.}\
  \bibnamefont {Yao}}, \bibinfo {author} {\bibfnamefont {P.}~\bibnamefont
  {Yeh}}, \bibinfo {author} {\bibfnamefont {A.}~\bibnamefont {Zalcman}}, \ and\
  \bibinfo {author} {\bibfnamefont {S.}~\bibnamefont {Zanker}},\ }\href@noop {}
  {\enquote {\bibinfo {title} {Observation of separated dynamics of charge and
  spin in the fermi-hubbard model},}\ } (\bibinfo {year} {2020}),\ \Eprint
  {http://arxiv.org/abs/2010.07965} {arXiv:2010.07965 [quant-ph]} \BibitemShut
  {NoStop}%
\bibitem [{\citenamefont {Feist}\ and\ \citenamefont
  {Garcia-Vidal}(2015)}]{PhysRevLett.114.196402}%
  \BibitemOpen
  \bibfield  {author} {\bibinfo {author} {\bibfnamefont {J.}~\bibnamefont
  {Feist}}\ and\ \bibinfo {author} {\bibfnamefont {F.~J.}\ \bibnamefont
  {Garcia-Vidal}},\ }\href {\doibase 10.1103/PhysRevLett.114.196402} {\bibfield
   {journal} {\bibinfo  {journal} {Phys. Rev. Lett.}\ }\textbf {\bibinfo
  {volume} {114}},\ \bibinfo {pages} {196402} (\bibinfo {year}
  {2015})}\BibitemShut {NoStop}%
\bibitem [{\citenamefont {Schachenmayer}\ \emph {et~al.}(2015)\citenamefont
  {Schachenmayer}, \citenamefont {Genes}, \citenamefont {Tignone},\ and\
  \citenamefont {Pupillo}}]{PhysRevLett.114.196403}%
  \BibitemOpen
  \bibfield  {author} {\bibinfo {author} {\bibfnamefont {J.}~\bibnamefont
  {Schachenmayer}}, \bibinfo {author} {\bibfnamefont {C.}~\bibnamefont
  {Genes}}, \bibinfo {author} {\bibfnamefont {E.}~\bibnamefont {Tignone}}, \
  and\ \bibinfo {author} {\bibfnamefont {G.}~\bibnamefont {Pupillo}},\ }\href
  {\doibase 10.1103/PhysRevLett.114.196403} {\bibfield  {journal} {\bibinfo
  {journal} {Phys. Rev. Lett.}\ }\textbf {\bibinfo {volume} {114}},\ \bibinfo
  {pages} {196403} (\bibinfo {year} {2015})}\BibitemShut {NoStop}%
\bibitem [{\citenamefont {Botzung}\ \emph {et~al.}(2020)\citenamefont
  {Botzung}, \citenamefont {Hagenm\"uller}, \citenamefont {Sch\"utz},
  \citenamefont {Dubail}, \citenamefont {Pupillo},\ and\ \citenamefont
  {Schachenmayer}}]{PhysRevB.102.144202}%
  \BibitemOpen
  \bibfield  {author} {\bibinfo {author} {\bibfnamefont {T.}~\bibnamefont
  {Botzung}}, \bibinfo {author} {\bibfnamefont {D.}~\bibnamefont
  {Hagenm\"uller}}, \bibinfo {author} {\bibfnamefont {S.}~\bibnamefont
  {Sch\"utz}}, \bibinfo {author} {\bibfnamefont {J.}~\bibnamefont {Dubail}},
  \bibinfo {author} {\bibfnamefont {G.}~\bibnamefont {Pupillo}}, \ and\
  \bibinfo {author} {\bibfnamefont {J.}~\bibnamefont {Schachenmayer}},\ }\href
  {\doibase 10.1103/PhysRevB.102.144202} {\bibfield  {journal} {\bibinfo
  {journal} {Phys. Rev. B}\ }\textbf {\bibinfo {volume} {102}},\ \bibinfo
  {pages} {144202} (\bibinfo {year} {2020})}\BibitemShut {NoStop}%
\bibitem [{\citenamefont {Berke}\ \emph {et~al.}(2020)\citenamefont {Berke},
  \citenamefont {Varvelis}, \citenamefont {Trebst}, \citenamefont {Altland},\
  and\ \citenamefont {DiVincenzo}}]{berke2020transmon}%
  \BibitemOpen
  \bibfield  {author} {\bibinfo {author} {\bibfnamefont {C.}~\bibnamefont
  {Berke}}, \bibinfo {author} {\bibfnamefont {E.}~\bibnamefont {Varvelis}},
  \bibinfo {author} {\bibfnamefont {S.}~\bibnamefont {Trebst}}, \bibinfo
  {author} {\bibfnamefont {A.}~\bibnamefont {Altland}}, \ and\ \bibinfo
  {author} {\bibfnamefont {D.~P.}\ \bibnamefont {DiVincenzo}},\ }\href@noop {}
  {\bibfield  {journal} {\bibinfo  {journal} {arXiv preprint arXiv:2012.05923}\
  } (\bibinfo {year} {2020})}\BibitemShut {NoStop}%
\bibitem [{\citenamefont {Temnov}\ and\ \citenamefont
  {Woggon}(2005)}]{PhysRevLett.95.243602}%
  \BibitemOpen
  \bibfield  {author} {\bibinfo {author} {\bibfnamefont {V.~V.}\ \bibnamefont
  {Temnov}}\ and\ \bibinfo {author} {\bibfnamefont {U.}~\bibnamefont
  {Woggon}},\ }\href {\doibase 10.1103/PhysRevLett.95.243602} {\bibfield
  {journal} {\bibinfo  {journal} {Phys. Rev. Lett.}\ }\textbf {\bibinfo
  {volume} {95}},\ \bibinfo {pages} {243602} (\bibinfo {year}
  {2005})}\BibitemShut {NoStop}%
\bibitem [{\citenamefont {Diniz}\ \emph {et~al.}(2011)\citenamefont {Diniz},
  \citenamefont {Portolan}, \citenamefont {Ferreira}, \citenamefont {G\'erard},
  \citenamefont {Bertet},\ and\ \citenamefont
  {Auff\`eves}}]{PhysRevA.84.063810}%
  \BibitemOpen
  \bibfield  {author} {\bibinfo {author} {\bibfnamefont {I.}~\bibnamefont
  {Diniz}}, \bibinfo {author} {\bibfnamefont {S.}~\bibnamefont {Portolan}},
  \bibinfo {author} {\bibfnamefont {R.}~\bibnamefont {Ferreira}}, \bibinfo
  {author} {\bibfnamefont {J.~M.}\ \bibnamefont {G\'erard}}, \bibinfo {author}
  {\bibfnamefont {P.}~\bibnamefont {Bertet}}, \ and\ \bibinfo {author}
  {\bibfnamefont {A.}~\bibnamefont {Auff\`eves}},\ }\href {\doibase
  10.1103/PhysRevA.84.063810} {\bibfield  {journal} {\bibinfo  {journal} {Phys.
  Rev. A}\ }\textbf {\bibinfo {volume} {84}},\ \bibinfo {pages} {063810}
  (\bibinfo {year} {2011})}\BibitemShut {NoStop}%
\bibitem [{\citenamefont {Agarwal}(1984)}]{PhysRevLett.53.1732}%
  \BibitemOpen
  \bibfield  {author} {\bibinfo {author} {\bibfnamefont {G.~S.}\ \bibnamefont
  {Agarwal}},\ }\href {\doibase 10.1103/PhysRevLett.53.1732} {\bibfield
  {journal} {\bibinfo  {journal} {Phys. Rev. Lett.}\ }\textbf {\bibinfo
  {volume} {53}},\ \bibinfo {pages} {1732} (\bibinfo {year}
  {1984})}\BibitemShut {NoStop}%
\bibitem [{\citenamefont {Yang}\ \emph {et~al.}(2020)\citenamefont {Yang},
  \citenamefont {Brehm}, \citenamefont {Lepp\"akangas}, \citenamefont {Guo},
  \citenamefont {Marthaler}, \citenamefont {Boventer}, \citenamefont {Stehli},
  \citenamefont {Wolz}, \citenamefont {Ustinov},\ and\ \citenamefont
  {Weides}}]{Yang-2020}%
  \BibitemOpen
  \bibfield  {author} {\bibinfo {author} {\bibfnamefont {P.}~\bibnamefont
  {Yang}}, \bibinfo {author} {\bibfnamefont {J.~D.}\ \bibnamefont {Brehm}},
  \bibinfo {author} {\bibfnamefont {J.}~\bibnamefont {Lepp\"akangas}}, \bibinfo
  {author} {\bibfnamefont {L.}~\bibnamefont {Guo}}, \bibinfo {author}
  {\bibfnamefont {M.}~\bibnamefont {Marthaler}}, \bibinfo {author}
  {\bibfnamefont {I.}~\bibnamefont {Boventer}}, \bibinfo {author}
  {\bibfnamefont {A.}~\bibnamefont {Stehli}}, \bibinfo {author} {\bibfnamefont
  {T.}~\bibnamefont {Wolz}}, \bibinfo {author} {\bibfnamefont {A.~V.}\
  \bibnamefont {Ustinov}}, \ and\ \bibinfo {author} {\bibfnamefont
  {M.}~\bibnamefont {Weides}},\ }\href {\doibase
  10.1103/PhysRevApplied.14.024025} {\bibfield  {journal} {\bibinfo  {journal}
  {Phys. Rev. Applied}\ }\textbf {\bibinfo {volume} {14}},\ \bibinfo {pages}
  {024025} (\bibinfo {year} {2020})}\BibitemShut {NoStop}%
\bibitem [{\citenamefont {Rodionov}\ \emph {et~al.}(2019)\citenamefont
  {Rodionov}, \citenamefont {Baburin}, \citenamefont {Gabidullin},
  \citenamefont {Maklakov}, \citenamefont {Peters}, \citenamefont {Ryzhikov},\
  and\ \citenamefont {Andriyash}}]{Rodionov2019ER}%
  \BibitemOpen
  \bibfield  {author} {\bibinfo {author} {\bibfnamefont {I.~A.}\ \bibnamefont
  {Rodionov}}, \bibinfo {author} {\bibfnamefont {A.~S.}\ \bibnamefont
  {Baburin}}, \bibinfo {author} {\bibfnamefont {A.~R.}\ \bibnamefont
  {Gabidullin}}, \bibinfo {author} {\bibfnamefont {S.~S.}\ \bibnamefont
  {Maklakov}}, \bibinfo {author} {\bibfnamefont {S.}~\bibnamefont {Peters}},
  \bibinfo {author} {\bibfnamefont {I.~A.}\ \bibnamefont {Ryzhikov}}, \ and\
  \bibinfo {author} {\bibfnamefont {A.~V.}\ \bibnamefont {Andriyash}},\ }\href
  {\doibase 10.1038/s41598-019-48508-3} {\bibfield  {journal} {\bibinfo
  {journal} {Scientific Reports}\ }\textbf {\bibinfo {volume} {9}},\ \bibinfo
  {pages} {12232} (\bibinfo {year} {2019})}\BibitemShut {NoStop}%
\bibitem [{\citenamefont {Chen}\ \emph
  {et~al.}(2014{\natexlab{b}})\citenamefont {Chen}, \citenamefont {Megrant},
  \citenamefont {Kelly}, \citenamefont {Barends}, \citenamefont {Bochmann},
  \citenamefont {Chen}, \citenamefont {Chiaro}, \citenamefont {Dunsworth},
  \citenamefont {Jeffrey}, \citenamefont {Mutus}, \citenamefont {O'Malley},
  \citenamefont {Neill}, \citenamefont {Roushan}, \citenamefont {Sank},
  \citenamefont {Vainsencher}, \citenamefont {Wenner}, \citenamefont {White},
  \citenamefont {Cleland},\ and\ \citenamefont
  {Martinis}}]{doi:10.1063/1.4863745}%
  \BibitemOpen
  \bibfield  {author} {\bibinfo {author} {\bibfnamefont {Z.}~\bibnamefont
  {Chen}}, \bibinfo {author} {\bibfnamefont {A.}~\bibnamefont {Megrant}},
  \bibinfo {author} {\bibfnamefont {J.}~\bibnamefont {Kelly}}, \bibinfo
  {author} {\bibfnamefont {R.}~\bibnamefont {Barends}}, \bibinfo {author}
  {\bibfnamefont {J.}~\bibnamefont {Bochmann}}, \bibinfo {author}
  {\bibfnamefont {Y.}~\bibnamefont {Chen}}, \bibinfo {author} {\bibfnamefont
  {B.}~\bibnamefont {Chiaro}}, \bibinfo {author} {\bibfnamefont
  {A.}~\bibnamefont {Dunsworth}}, \bibinfo {author} {\bibfnamefont
  {E.}~\bibnamefont {Jeffrey}}, \bibinfo {author} {\bibfnamefont {J.~Y.}\
  \bibnamefont {Mutus}}, \bibinfo {author} {\bibfnamefont {P.~J.~J.}\
  \bibnamefont {O'Malley}}, \bibinfo {author} {\bibfnamefont {C.}~\bibnamefont
  {Neill}}, \bibinfo {author} {\bibfnamefont {P.}~\bibnamefont {Roushan}},
  \bibinfo {author} {\bibfnamefont {D.}~\bibnamefont {Sank}}, \bibinfo {author}
  {\bibfnamefont {A.}~\bibnamefont {Vainsencher}}, \bibinfo {author}
  {\bibfnamefont {J.}~\bibnamefont {Wenner}}, \bibinfo {author} {\bibfnamefont
  {T.~C.}\ \bibnamefont {White}}, \bibinfo {author} {\bibfnamefont {A.~N.}\
  \bibnamefont {Cleland}}, \ and\ \bibinfo {author} {\bibfnamefont {J.~M.}\
  \bibnamefont {Martinis}},\ }\href {\doibase 10.1063/1.4863745} {\bibfield
  {journal} {\bibinfo  {journal} {Applied Physics Letters}\ }\textbf {\bibinfo
  {volume} {104}},\ \bibinfo {pages} {052602} (\bibinfo {year}
  {2014}{\natexlab{b}})}\BibitemShut {NoStop}%
\end{thebibliography}

%

\end{document}